\documentclass[12pt]{article}

\usepackage{graphicx}
\usepackage{hyperref}
\usepackage{longtable}
\usepackage{booktabs}
\usepackage{float}
\usepackage{lineno}
\usepackage{hyperref}
\hypersetup{colorlinks=true,citecolor=blue,linkcolor=blue,urlcolor=blue,breaklinks=true}
\RequirePackage{doi}

\begin{document}

\title{\bf Spin state and moment of inertia of Venus}

\author{Jean-Luc Margot$^{1,2}$\footnote{email: jlm@epss.ucla.edu}\,,
      Donald B. Campbell$^{3}$,  Jon D. Giorgini$^{4}$,\\
      Joseph S. Jao$^{4}$, Lawrence G. Snedeker$^{4}$, Frank D. Ghigo$^{5}$, Amber Bonsall$^{5}$ \\ \\
\normalsize{$^{1}$Department of Earth, Planetary, \& Space Sciences, UCLA, Los Angeles, CA 90095, USA,}\\
\normalsize{$^{2}$Department of Physics \& Astronomy, UCLA, Los Angeles, CA 90095, USA,}\\
\normalsize{$^{3}$Department of Astronomy, Cornell University, Ithaca, NY 14853, USA,}\\
\normalsize{$^{4}$Jet Propulsion Laboratory, Pasadena, CA 91109, USA,}\\
\normalsize{$^{5}$National Radio Astronomy Laboratory, Green Bank, WV 24944}\\
}

\date{}

\maketitle

{\bf Fundamental properties of the planet Venus, such as its internal mass distribution and variations in length of day, have remained unknown.  We used Earth-based observations of radar speckles tied to the rotation of Venus obtained in 2006--2020 to measure its spin axis orientation, spin precession rate, moment of inertia, and length-of-day variations.  Venus is tilted by 2.6392 $\pm$ 0.0008
degrees ($1\sigma$) with respect to its orbital plane.  The spin axis precesses at a rate of 44.58 $\pm$ 3.3 arcseconds per year ($1\sigma$), which gives a normalized moment of inertia of 0.337 $\pm$ 0.024 and yields a rough estimate of the size of the core.  The average sidereal day on Venus in the 2006--2020 interval is 243.0226 $\pm$ 0.0013 Earth days ($1\sigma$).  The spin period of the solid planet exhibits variations of 61 ppm ($\sim$20 minutes) with a possible diurnal or semidiurnal forcing.  The length-of-day variations imply that changes in atmospheric angular momentum of at least $\sim$4\% are transferred to the solid planet.}


Venus is Earth’s nearest planetary neighbor and closest analog in the
Solar System in terms of mass, radius, and density. However, Venus
remains enigmatic on a variety of fundamental levels: the size of its
core is unknown \cite{smre18}; whether the core is solid or liquid is
uncertain \cite{dumo16, orou18}; 
and estimates of its average
spin period are discordant
\cite{davi92,muel12,camp19}.
Venus is also distinctive because of its 243-day retrograde rotation
and 4-day atmospheric superrotation,
neither of which is fully understood
\cite{yode97,sanc17,hori20}.
High-precision measurements of the spin state enable progress in all
these areas.

The polar moment of inertia provides an integral constraint on the
distribution of mass in a planetary interior, $C = \int_V \rho r^2
dV$, where the volume integral includes the mass density $\rho$ at
each point and the square of the distance $r$ from the spin axis.
Along with bulk density, the moment of inertia is arguably the most
important quantity needed to determine the internal structure of a
planetary body.  In particular, it can be used to
place bounds on the size
of the core, which is essential in understanding a planet's thermal,
spin, and magnetic evolutionary histories.
Seismology, which has been conducted on Earth, Moon, and
Mars, provides a powerful probe of planetary interiors, but is
considered ``a distant goal'' at Venus \cite{vexag19} due to the
planet's extreme surface temperature ($\sim$740 K) and pressure
($\sim$90 atm).

Gravitational torques from the Sun result in a precession of the spin
axis, which is similar to the motion of a spinning top.  The rate of
precession is inversely proportional to the polar moment of
inertia \cite{kaul68short}:
\begin{equation}
  \frac{d\psi }{dt}=
  \frac{3}{2}\left( \frac{n^{2}}{\omega }\right)  J_2  \left(\frac{MR^2}{C} \right) \cos{\theta}, 
\label{eq-precess}
\end{equation}
where $n$ is the orbital mean motion, $\omega$ is the spin rate,
$J_2$ is the second-degree coefficient in the spherical harmonic
expansion of the gravity field, $M$ is the mass, $R$ is the radius,
and $\theta$ is the obliquity, or angular separation between spin and
orbit poles.  Measurements of the spin precession rates of Earth
(50.2877 arcsec/y) and Mars (7.576 arcsec/y) yield $C/MR^2$=0.3307
\cite{will94} and $C/MR^2$=0.3662 \cite{folk97}, respectively.
The predicted precession rate of Venus for a nominal $C/MR^2=0.336$ is
$44.75$ arcsec/y,
where we have used $n=585.17$~deg/y \cite{stan13},
$\omega=541.06$~deg/y,
$J_2=4.40454 \times 10^{-6}$
\cite{kono99}, and $\theta = 2.639$~deg.  This value
is in good agreement with a previous estimate \cite{cott09} and implies a
precession cycle of $\sim$29,000~y.

Although the predicted precession rate of Venus is similar to that of
Earth, the motion of the spin pole in inertial space is only
2.06~arcsec/y because of Venus's small obliquity.  Detection of the
precession was out of reach of the 1990--1994 Magellan spacecraft
mission despite its
extensive radar coverage with resolution as fine
as 100 m \cite{saun92}.  The best Magellan estimates of the spin axis
orientation from analysis of radar and gravity data have uncertainties
of 46~arcsec \cite{davi92} and 14 arcsec \cite{kono99}, respectively.
If a future mission were to measure the spin axis
orientation with infinite precision at an epoch circa 2040,
the measured precession excursion
of $\sim$100 arcsec since the Magellan epoch would be determined with
14 arcsec errors at best, or 14\% uncertainties, which is not
geophysically useful.
Likewise, if a future orbiter with 5-year duration were to measure the inertial positions of landmarks with 30~m precision, it would detect the $\sim$10 arcsec precession over the mission duration, which corresponds to a maximum displacement of $\sim$300 m, with 10\% uncertainties at best.
A useful measurement could be obtained with
telemetry data from multiple landers, as in the case of Mars,
but the technical challenge and cost of this endeavor make it
improbable in the foreseeable future.

Measurements of the average spin period of Venus with $<$1\% precision
were first obtained by tracking the positions of surface features
detectable in Earth-based radar images spanning multiple conjunctions
\cite{shap79,zoha80,shap90,slad90}.
The most precise value to date was obtained by analyzing the positions
of hundreds of landmarks detectable in Magellan spacecraft radar
images
recorded in the early 1990s, which resulted in a $\sim$500-day-average
spin period estimate of 243.0185 $\pm$ 0.0001
days~\cite{davi92}.  Certain features observed in Magellan
images were also detected in Venus Express spacecraft images obtained
circa 2007, enabling a $\sim$16-year-average spin period estimate of
243.023 $\pm$ 0.001 days~\cite{muel12}.
Recent measurements of the positions of surface features in
Earth-based radar images obtained between 1988 and 2017 yielded a
$\sim$29-year-average spin period estimate of 243.0212 $\pm$ 0.0006
days \cite{camp19}.  None of these estimates were of sufficient
precision to detect either sidereal length-of-day (LOD) fluctuations or the
precession of the spin axis.

The maintenance of a 243-day retrograde spin requires explanation
because solid body tides raised by the Sun would synchronize the spin
of Venus to its 225-day orbital period in the absence of other
forces~\cite{gold66venus,goldpeal67}. Gold and Soter
\cite{gold69venus} and others \cite{inge78,dobr80,corr01} proposed that solar
torques on a thermally-induced atmospheric tide might counteract the
solid-body torques and stabilize the spin.
In this process, atmospheric mass decreases on the hot afternoon side
of the planet and increases on the colder morning side.  This
imbalance creates an atmospheric bulge that leads the sub-solar point,
whereas the bulge due to tides raised on the solid body lags the sub-solar point.  The
opposing solar torques may therefore stabilize the spin rate.  The
solid-body torque is relatively independent of the spin rate, but the
strength of the torque on the atmospheric bulge has a semidiurnal
dependence~\cite{gold69venus,inge78,dobr80}.  It is thought that the
spin rate settles where the two torques balance each other.  However,
the magnitude of variations around the equilibrium and the nature of
the response to departures from equilibrium have not been elucidated.

With a spin rate controlled by a thermally driven atmospheric tide,
the rotation of the planet likely
changes as a result of variations in albedo \cite{yode97}, orbital
eccentricity \cite{bill05venus}, insolation, climate, and weather.
The insolation changes by 3\% as Venus revolves around the Sun with
its current eccentricity of 0.007 and by 15\% with the long-term
average eccentricity of 0.035 \cite{bill05venus}.
The planet may therefore exhibit daily and seasonal fluctuations in
LOD superposed on a complex spin rate evolution on longer timescales.

Atmospheric angular momentum (AAM) on Earth varies by tens of percent
\cite{hide80} and results in LOD variations on the order of
milliseconds.
With its massive atmosphere, Venus has an estimated AAM value ($L \sim
2.9 \times 10^{28}$ kg~m$^2$~s$^{-1} \pm30\%$) \cite{sanc17}
that is
$\sim$180 times larger than Earth's
and the
atmospheric fraction of total planetary angular momentum is
$\sim$60,000
larger than Earth's.
If a fraction $\epsilon$ of
AAM is transferred to the solid
planet, the rotation period $P$ changes by
$\Delta P / P = -\epsilon L / C\omega$.
For Venus, $\Delta P \simeq -9.4 \epsilon$ h.
Peak-to-peak estimates of AAM-induced LOD variations based on Global
Circulation Model (GCM) simulations currently span at least two orders
of magnitude, with values of $\epsilon$ ranging from below $\sim$0.1\%
\cite{lebo10,cott11} on diurnal timescales to above
$\sim$15\%~\cite{pari11} on decadal timescales.

\paragraph*{Observations}
We obtained high-precision measurements of the instantaneous spin state of Venus
with a radar speckle tracking technique that requires
two telescopes and does not involve imaging
(see `Radar speckle tracking' in Methods).
We used the
70 m antenna (DSS-14) at Goldstone, California
(35.24$^\circ$N, -116.89$^\circ$E) and transmitted a circularly
polarized, monochromatic signal at a frequency of 8560~MHz ($\lambda =
3.5$ cm) and power of $\sim$200--400~kW.  Radar echoes were recorded at
DSS-14 and also at the
100 m
Green Bank Telescope (GBT) in West
Virginia (38.24$^\circ$N, -79.84$^\circ$E)
with fast sampling systems~\cite{pfs}.

\begin{figure}[p]
	\centering
	\includegraphics[angle=0,width=5in]{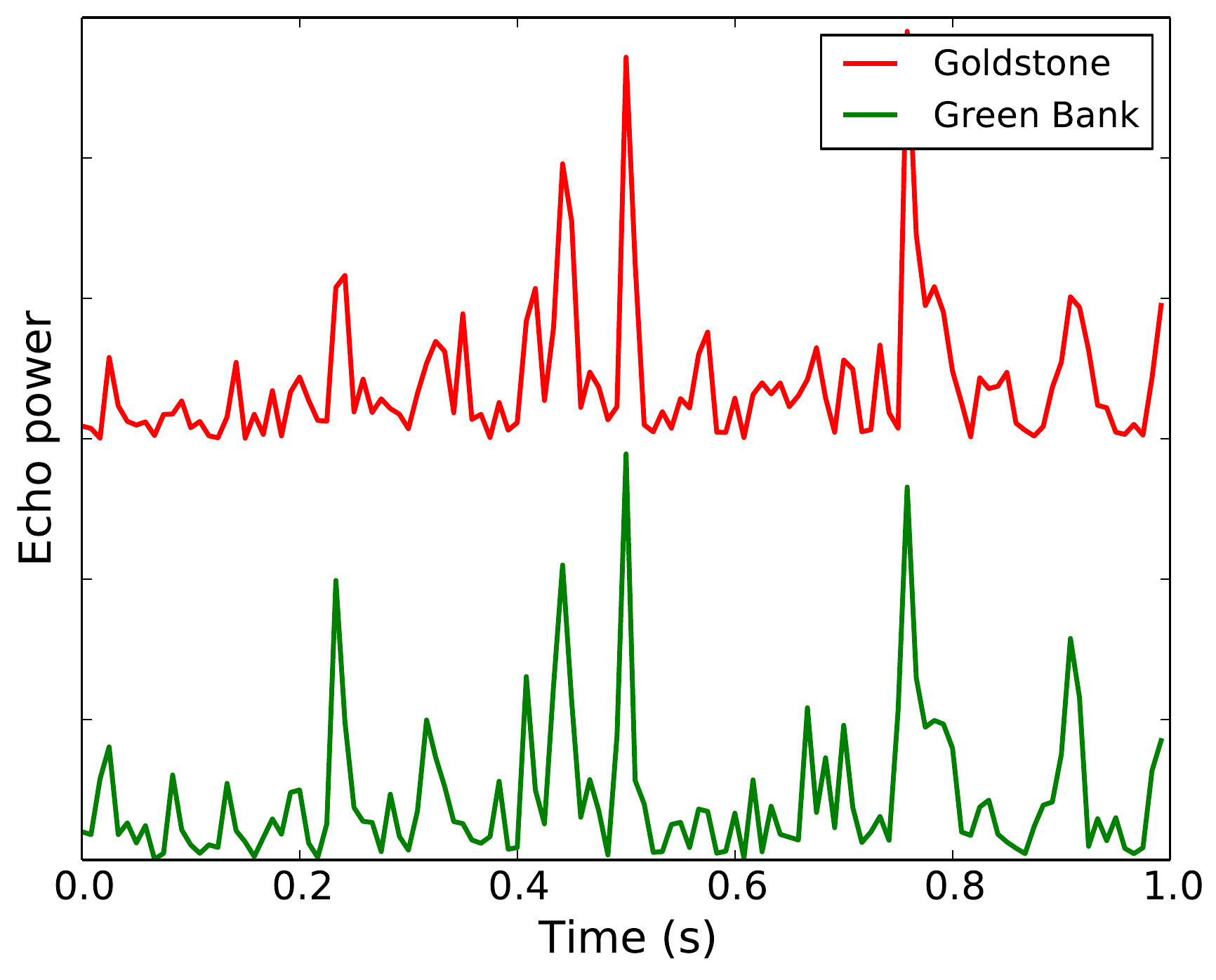}
	\caption{
          Representative variations in radar echo power ({\em speckles}) from
          observations of Venus with the Goldstone Solar System Radar
          and Green Bank Telescope at 8560 MHz on 2016 Nov.~26.  The
          GBT echo was shifted in time by 20~s to illustrate the high
          degree of correlation between the received waveforms when
          the speckle trajectory is aligned with the antenna baseline.}
\label{fig-speckles} 
\end{figure}

\begin{figure}[p]
	\centering
	\includegraphics[angle=0,width=6.5in]{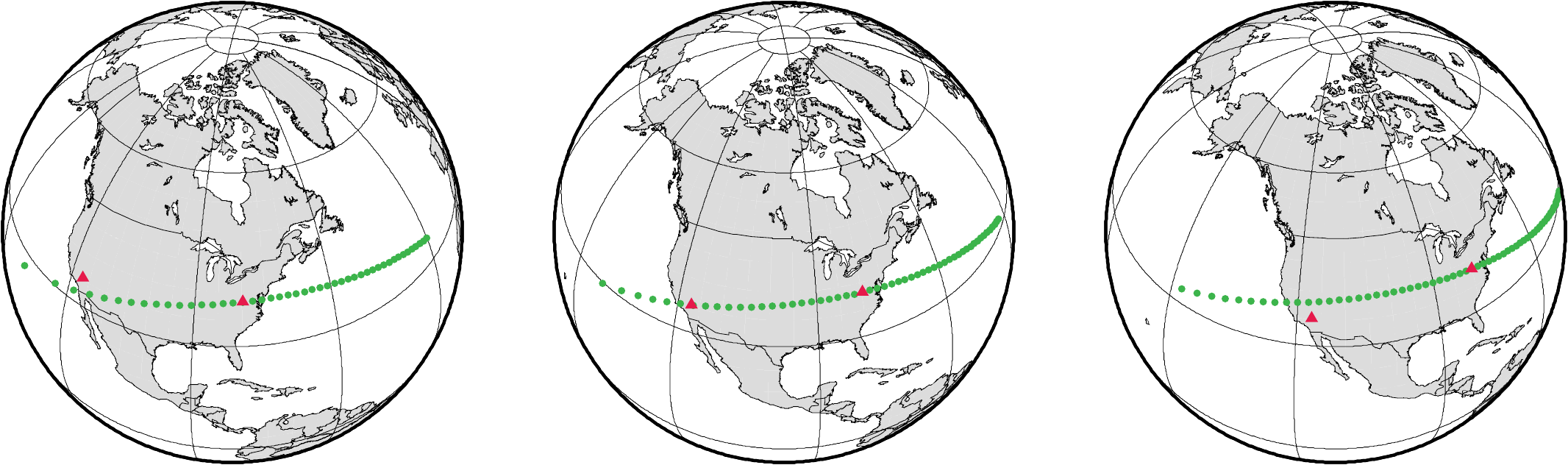}
	\caption{
          Radar echoes from Venus sweep over the surface of
          the Earth during the 2020 Sept.\ 08 observations.  Diagrams
          show the trajectory of the speckles one hour before (left),
          during (center), and one hour after (right) the epoch of
          maximum correlation.  Echoes from two receive stations (red
          triangles) exhibit a strong correlation when the antennas
          are suitably aligned with the trajectory of the speckles
          (green dots shown with a 1~s time interval).}
\label{fig-path} 
\end{figure}

\begin{figure}[h]
  \begin{center}
    \begin{tabular}[h]{cc}

      \includegraphics[angle=0,width=3in]{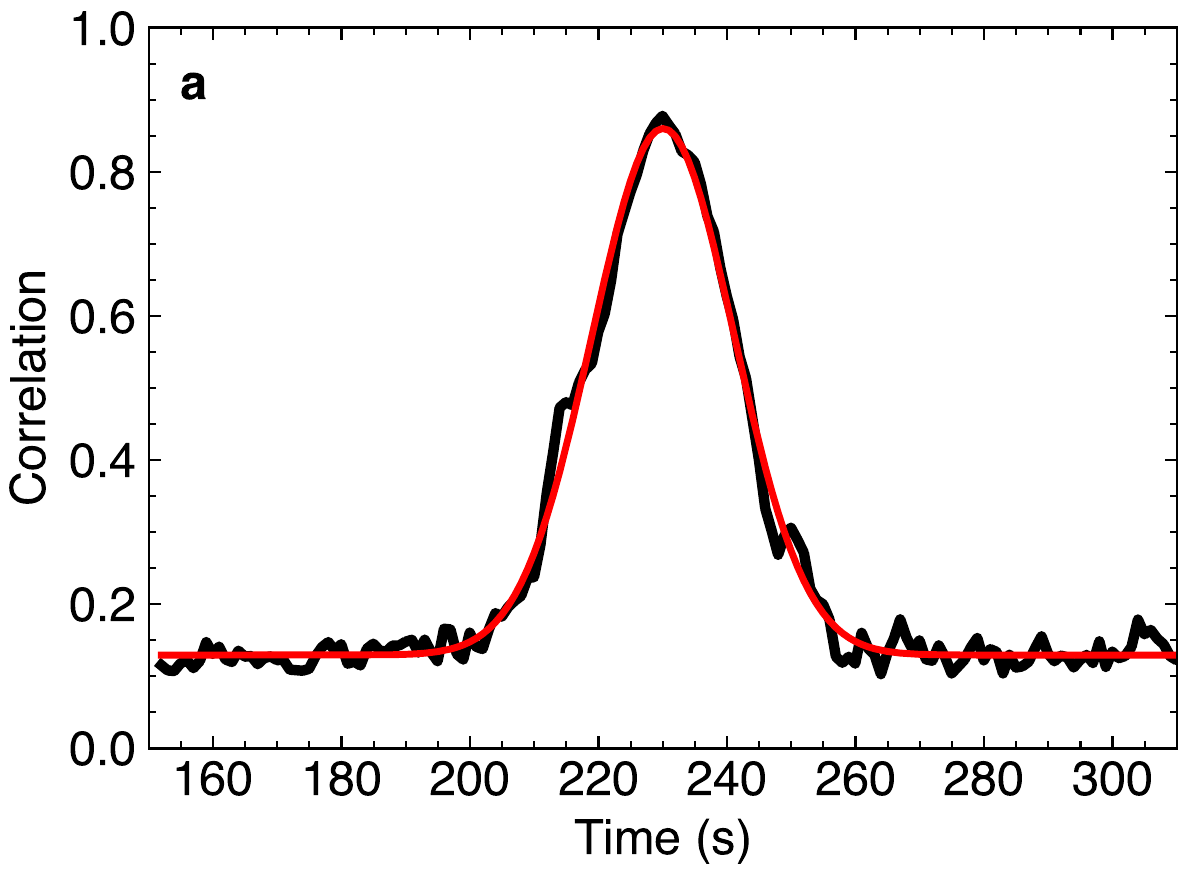} &
      \includegraphics[angle=0,width=3in]{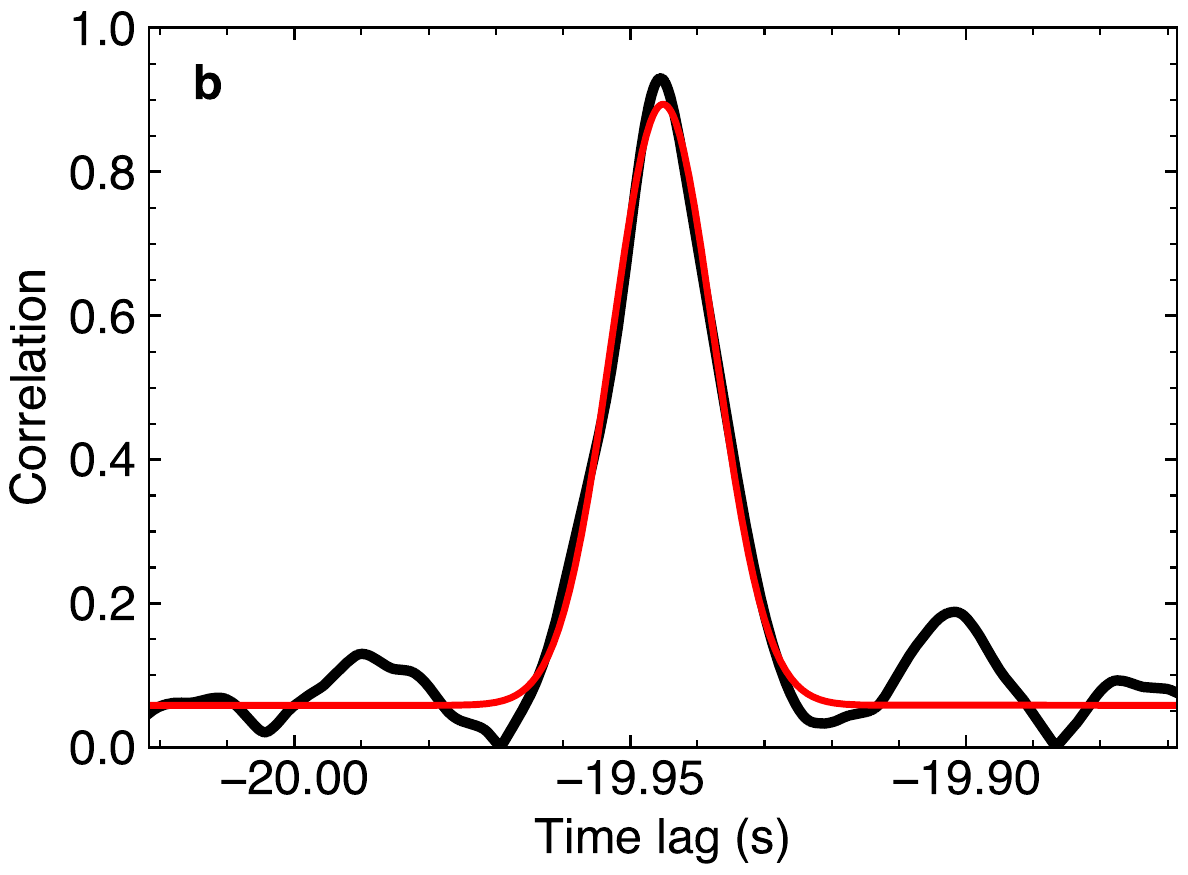} \\

    \end{tabular}
    \caption{Representative functions obtained by cross-correlating
      Venus radar echoes recorded at Goldstone, CA, and Green Bank,
      WV, on 26 Nov 2016, with Gaussian fits shown in red.  ({\bf a})
      Time evolution of the maximum in the cross-correlation function
      with echo time series decimated to a sampling rate of 200 Hz.
      The time of the correlation maximum represents the epoch at
      which the speckle trajectory is aligned with both antennas and is
      diagnostic of spin axis orientation.  ({\bf b})
      Cross-correlation of echo time series with 5000 Hz sampling rate
      and 5 ms moving average at the time of the correlation maximum
      shown in panel {\bf a}.  The time lag corresponding to the correlation
      maximum is diagnostic of the instantaneous spin rate.
      Successive measurements of this time lag yield the estimates and
      $\sim$5 ppm fractional uncertainties listed in Table~1.  }
    \label{fig-corr}
  \end{center}
\end{figure}

Radar echoes from solid surfaces are speckled.
The radar
speckles are tied to the rotation of Venus and sweep over the surface
of the Earth with a trajectory that is occasionally aligned with the
two telescopes
(Figs.\ \ref{fig-speckles} and \ref{fig-path}).
We cross-correlated the echo time series received at each telescope and
obtained strong correlations that lasted $\sim$30~s
(Fig.~\ref{fig-corr}).  The {\em epoch} at which the high correlation
occurs is diagnostic of the spin axis orientation.  The {\em time lag}
at which the
correlation maximizes yields a measurement of the instantaneous spin
period.

We attempted to observe Venus on 121 instances between 2006 and 2020
(Supplementary Table~\ref{tab-all}) and were successful on 21
occasions, with observing circumstances reported in Supplementary
Table~\ref{tab-geom}.
The observing protocol and data reduction technique
(see `Observing protocol' and `Data reduction technique' in Methods)
closely followed those used for
similar
measurements at Mercury \cite{marg07,marg12jgr}
that
were
confirmed at the 1\% level by subsequent spacecraft observations
\cite{star15grl}.

We fit Gaussians to the correlation functions in order to obtain
estimates of the epochs of correlation maximum $\hat t$.  We also
obtained estimates of the time lags $\hat\tau$ that maximize the
correlation functions (Fig.~\ref{fig-corr}, Table~\ref{tab-obs}).
With large signal-to-noise ratios, Gaussian centroid locations can be
determined with a precision that is a small fraction of the widths of
the correlation functions~\cite{bend86}.
Measurement residuals and spread in consecutive estimates suggest that
epochs of correlation maxima and time lags can be determined to
precisions of $\sim$0.3~s and $\sim$0.1~ms from initial widths of
$\sim$10~s and $\sim$7~ms, respectively.
We used the $\hat t$ and $\hat \tau$ observables to determine the instantaneous spin state of Venus
(See `Conversion of observables to spin state estimates in Methods).

\begin{table}[p]
\begin{tabular}{rrrrrrrrr}
{Date} &
{Epoch (MJD)} &
{$A$} &   
{$w$ (s)} &   
{Time lag (s)} & 
{$\epsilon$ (ppm)} &
{$P$ (days)} 
\\
\hline
060128  &  53763.69074757  &   0.593  &   4.70 &	 -149.224656 &  6.09 &	   243.01724  \\
060129  &  53764.69293247  &   0.622  &   5.11 &	 -142.602992 &  7.42 &	   243.02106  \\
060207  &  53773.69201430  &   0.573  &   6.24 &	  -95.544513 &  6.57 &	   243.01721  \\
060214  &  53780.67533999  &   0.594  &  12.52 &	  -74.199631 &  3.93 &	   243.02289  \\
060219  &  53785.65987182  &   0.696  &   9.66 &	  -63.745660 &  4.44 &	   243.02168  \\
090614  &  54996.76400084  &   0.709  &   9.27 &	  -20.401318 &  3.07 &	   243.01596  \\
090801  &  55044.63200521  &   0.583  &  13.98 &	  -17.883815 &  5.76 &	   243.01594  \\
120310  &  55996.01816745  &   0.693  &   4.85 &	  -18.228874 &  4.44 &	   243.02257  \\
120311  &  55997.01549785  &   0.615  &   5.32 &	  -18.748452 &  3.71 &	   243.02468  \\
120314  &  56000.00753916  &   0.654  &   5.28 &	  -20.327400 &  7.44 &	   243.02073  \\
120315  &  56001.00490798  &   0.663  &   5.65 &	  -20.860567 &  5.54 &	   243.01651  \\
140312  &  56728.58912128  &   0.705  &   9.73 &	  -36.100686 &  2.19 &	   243.02960  \\
140314  &  56730.58274410  &   0.764  &   9.81 &	  -34.453769 &  2.73 &	   243.02925  \\
140315  &  56731.57957689  &   0.807  &   9.63 &	  -33.652202 &  2.41 &	   243.03075  \\
161122  &  57714.87788177  &   0.693  &  12.90 &	  -19.417283 &  4.60 &	   243.02861  \\
161125  &  57717.87004808  &   0.760  &  11.75 &	  -19.821632 &  4.01 &	   243.02932  \\
161126  &  57718.86741525  &   0.779  &  11.45 &	  -19.945238 &  3.29 &	   243.02635  \\
190206  &  58520.68253779  &   0.562  &  14.52 &	  -21.925929 &  5.68 &	   243.02289  \\
190207  &  58521.67959325  &   0.564  &  14.83 &	  -21.846558 &  4.72 &	   243.02335  \\
190208  &  58522.67663675  &   0.568  &  12.70 &	  -21.759899 &  5.23 &	   243.02186  \\
200908  &  59100.52352634  &   0.497  &   8.02 &	  -17.088679 &  4.51 &	   243.01782  \\
\end{tabular}                                                                 
\caption{Measurements of the instantaneous spin state of Venus.  The epoch of correlation maximum is
  reported as a Modified Julian Date (MJD).  It is the centroid of a
  Gaussian of amplitude $A$ and standard deviation $w$. The time lag
  indicates the time interval for speckles to travel from one
  station to the other at the corresponding epoch.  The reference
  epochs correspond to arrival times at the GBT, and the negative lag
  values indicate that Venus speckles travel from west to east.  The
  $1\sigma$ fractional uncertainty $\epsilon$ that applies to both the time lag
  and spin period is empirically determined from successive
  measurements.  The last column indicates the instantaneous, sidereal
  spin period in Earth days, after application of small refraction
  corrections (Table~\ref{tab-geom}).
    }
    \label{tab-obs}
\end{table}

\paragraph*{Spin axis orientation, precession, and moment of inertia}

\begin{figure}[h]
  \centering
    \includegraphics[angle=0,width=5.1in]{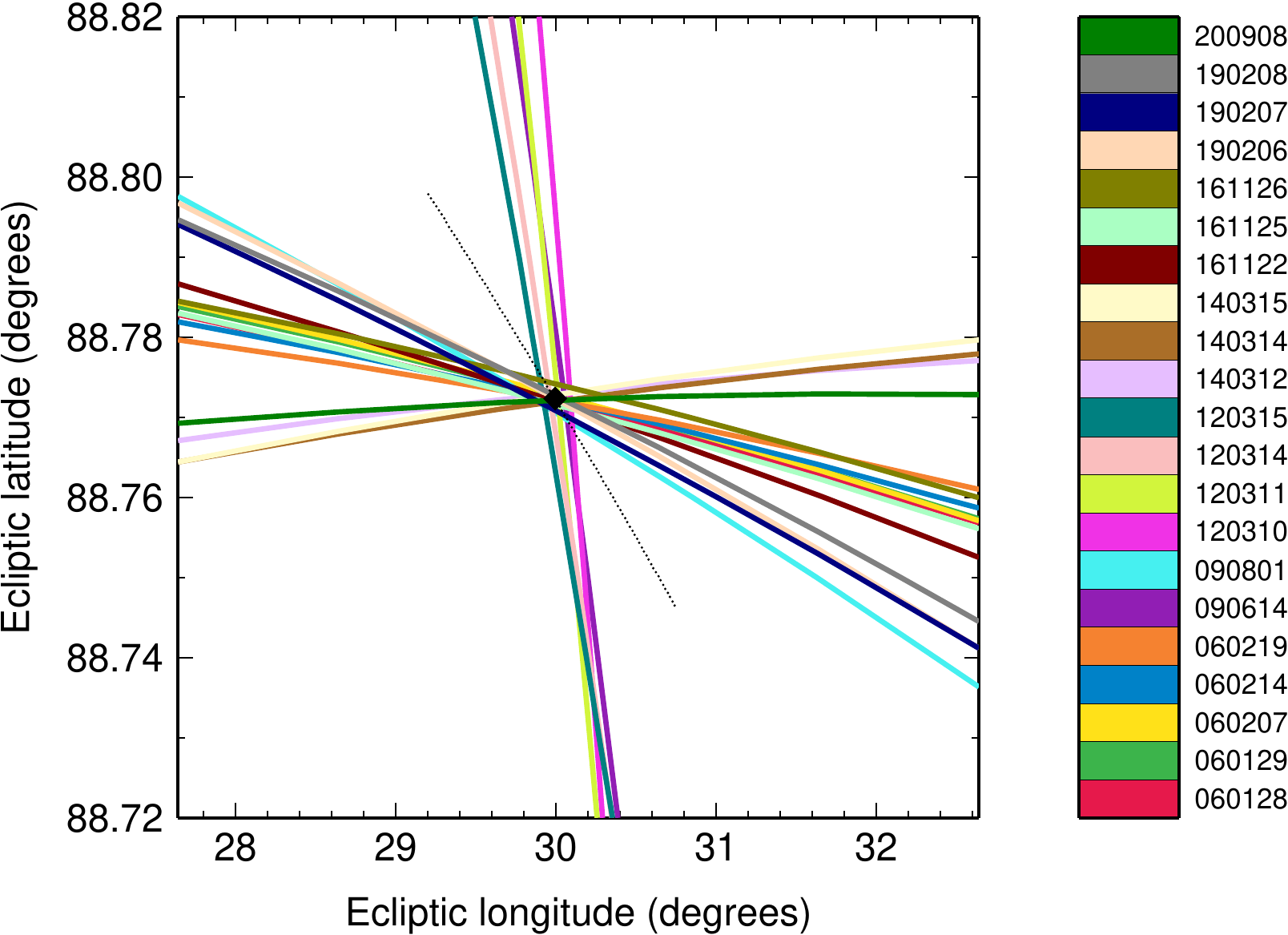} \\  
    \caption{
Constraints on the spin axis orientation of Venus
      obtained with Goldstone-GBT observations of radar speckles.
      Each colored line represents a
      measurement of the epoch of correlation maximum that traces a
      narrow error ellipse on the celestial sphere.  The orientation
      of each line is related to the ecliptic longitude of the
      projected baseline at the time of observations (Supplementary
      Table \ref{tab-geom}).  The best-fit spin axis orientation is
      shown by a diamond at the intersection of the colored lines.
      All measurements have been precessed to the J2000.0 epoch.  The
      black dotted line represents the trace of the spin axis
      orientation on the celestial sphere as a result of spin
      precession between 1950 and 2050.  }
  \label{fig-venspin}
\end{figure}

The velocity vector of the speckle pattern lies in the plane
perpendicular to the component of the spin vector that is
perpendicular to the line of sight.  The correlation of Goldstone and
GBT radar echoes is large only when the antenna separation vector
projected perpendicular to the line of sight, or {\em projected
  baseline}, lies in the same plane.  Correlation epoch measurements
provide tight constraints on the component of the spin that is
perpendicular to the line of sight and loose constraints on the
orthogonal component.
As a result, each epoch measurement delineates a narrow error ellipse
for the orientation of the spin axis on the celestial sphere.  We
obtained intersecting error ellipses by observing Venus at a variety
of orientations
(Fig.~\ref{fig-venspin}).

We used the epochs of correlation maxima (Table \ref{tab-obs}) with
uniform uncertainties in a three-parameter
least-squares fit to estimate the spin axis orientation of Venus as
well as its precession rate.  The first two adjustable parameters are
the right ascension (RA) and declination (DEC) of the spin axis in the
equatorial frame of J2000.0.  The third adjustable parameter is the
normalized moment of inertia $C/MR^2$, with precession modeled
according to equation (\ref{eq-precess}).  Post-fit residuals have a
standard deviation of 0.32 s and their distribution is unremarkable
(Supplementary Fig.~\ref{fig-residuals}).
We estimated confidence intervals with 2000 bootstrap trials, which
confirmed the robustness of the fit results with respect to inclusion
or exclusion of certain data points.  Results are listed in
Table~\ref{tab-res}.

\begin{table}[htb!]
  \begin{center}
\begin{tabular}{rrrr}
{Quantity} & 
{Least squares} &
{Bootstrap mean} &
{Std. dev.} 
\\
\hline       
RA (deg)  & 272.73911 & 272.73912 & 0.0008  \\
DEC (deg) &  67.15105 &  67.15100 & 0.0007  \\
$d\psi/dt$ ($^{\prime\prime}$/y) & -44.89  &  -44.58 & 3.3\\
$C/MR^2$  &    0.3350 &    0.3373 &  0.024  \\
$C$ (10$^{37}$ kg~m$^2$) & 5.972 & 6.013     &   0.43 \\
\end{tabular}
\caption{Estimates of Venus spin axis orientation, precession rate,
  and moment of inertia from least-squares fit and bootstrap analysis.
  Angles refer to epoch J2000.0.}
    \label{tab-res}
  \end{center}
  \end{table}

\begin{figure}[H]
  \centering
    \begin{tabular}[h]{cc}
  \includegraphics[width=3.2in]{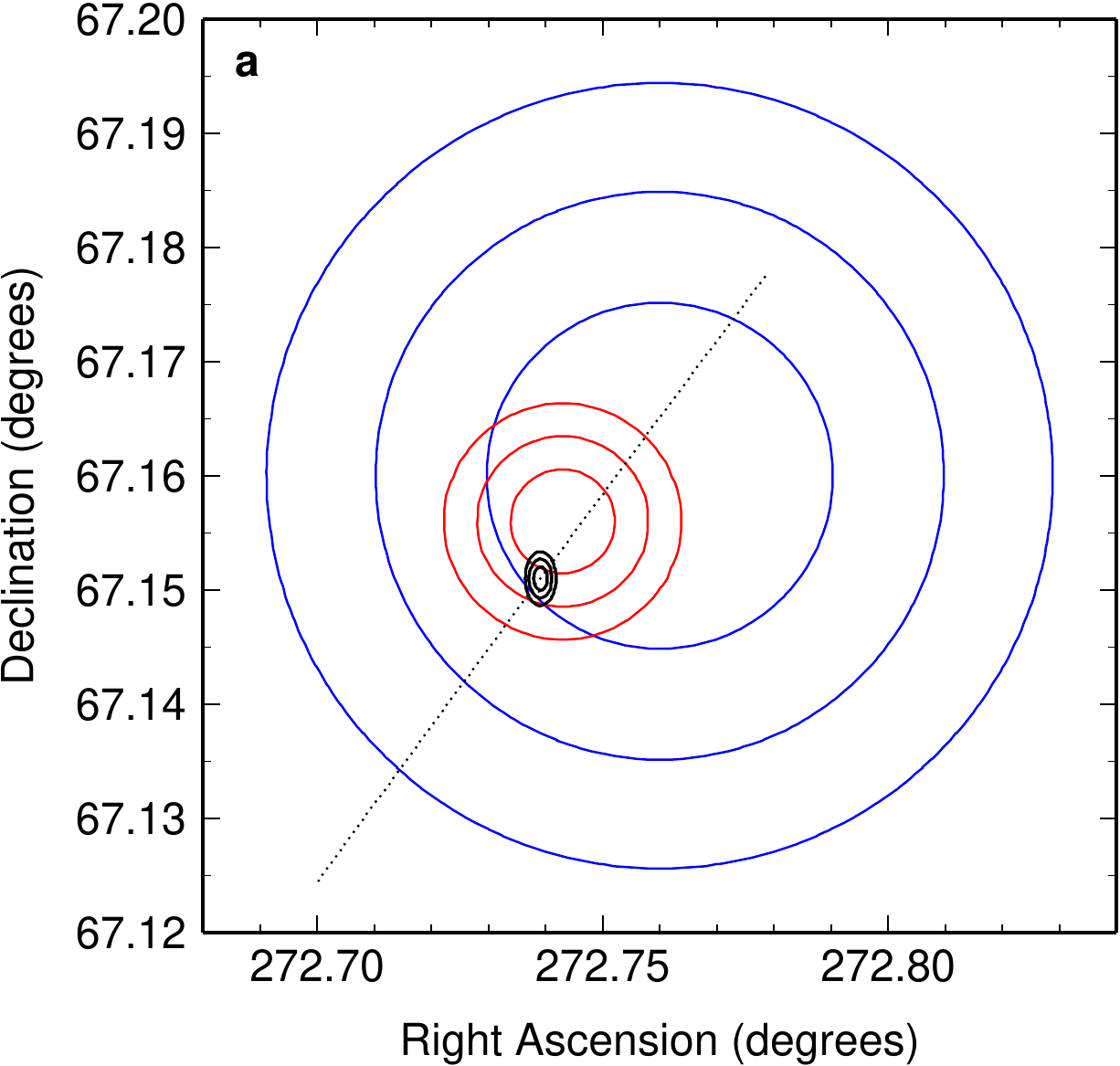} &
  \includegraphics[width=3.2in]{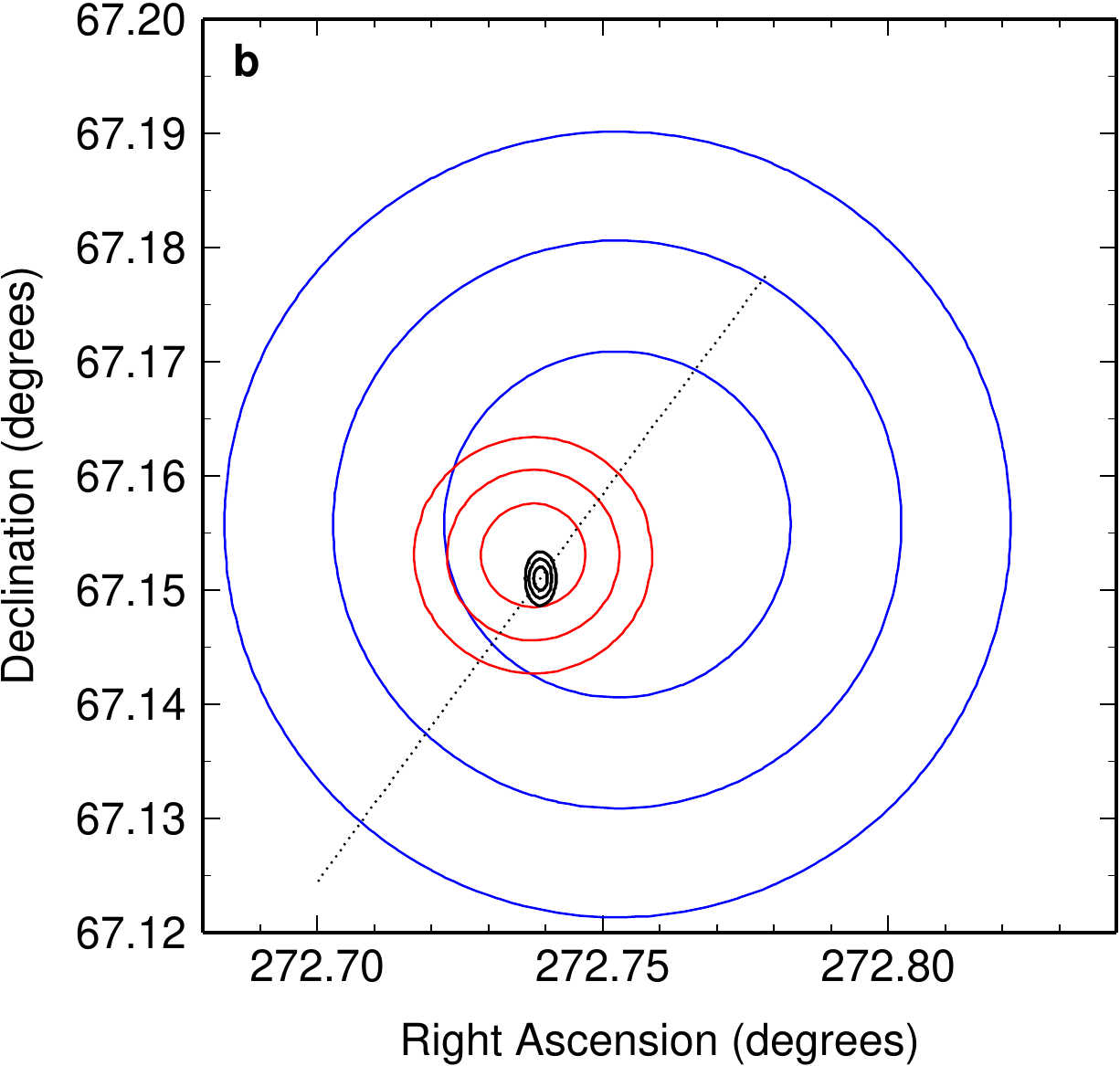} \\
    \end{tabular}
    \caption{ Spin axis orientation of Venus shown with
      1-, 2-, and
      3-sigma uncertainties (2D confidence intervals at 68.3\%,
      95.4\%, and 99.7\% levels, respectively).  The epoch J2000.0
      solution based on Goldstone-GBT observations is shown in black.
      The Magellan solutions of Davies et al.~\cite{davi92} and
      Konopliv et al.~\cite{kono99} are shown in blue and red,
      respectively.  The black dotted line represents the trace of the
      spin axis orientation on the celestial sphere as a result of
      spin precession between 1950 and 2050.  ({\bf a}) Magellan
      solutions as published by Davies et al.~\cite{davi92} and
      Konopliv et al.~\cite{kono99}.  ({\bf b}) Magellan solutions
      precessed to epoch J2000.0 from the mid-point of the corresponding
      observation intervals (Jan 1991 and Sep 1993, respectively).
    }
  \label{fig-magellan}
\end{figure}

The spin axis orientation of Venus is determined with an overall
precision of 2.7 arcsec, which improves upon the Magellan estimates by
a factor of 5--15 (Fig.~\ref{fig-magellan}).  At first glance, our
estimates are only marginally consistent with the Magellan estimates.
However, the spin axis orientation measured by Magellan in the early
1990s is not directly comparable to our solution, which has a 
reference epoch of J2000.0.  If we use our estimate of the precession
rate and precess the Magellan estimates to epoch J2000.0, we find that
our values fall well within the Magellan
$1\sigma$ uncertainty
contours (Fig.~\ref{fig-magellan}).


Our improved value of the obliquity of Venus is 2.6392 $\pm$ 0.0008
deg  ($1\sigma$), where we have used a recent determination of the
orbital plane orientation with RA=278.007642 deg and DEC=65.566999 deg \cite{stan13}.
The origin and maintenance of the obliquity has been linked to
planetary perturbations, core-mantle friction, and atmospheric torques
\cite{yode97,corr03a,corr03b}.  If the core is liquid, the obliquity
estimate
can be used to place bounds on the viscosity or
ellipticity of the core, provided that atmospheric torques are modeled
accurately \cite{yode97}.

\begin{figure}[H]
  \centering
    \includegraphics[angle=0,width=4in]{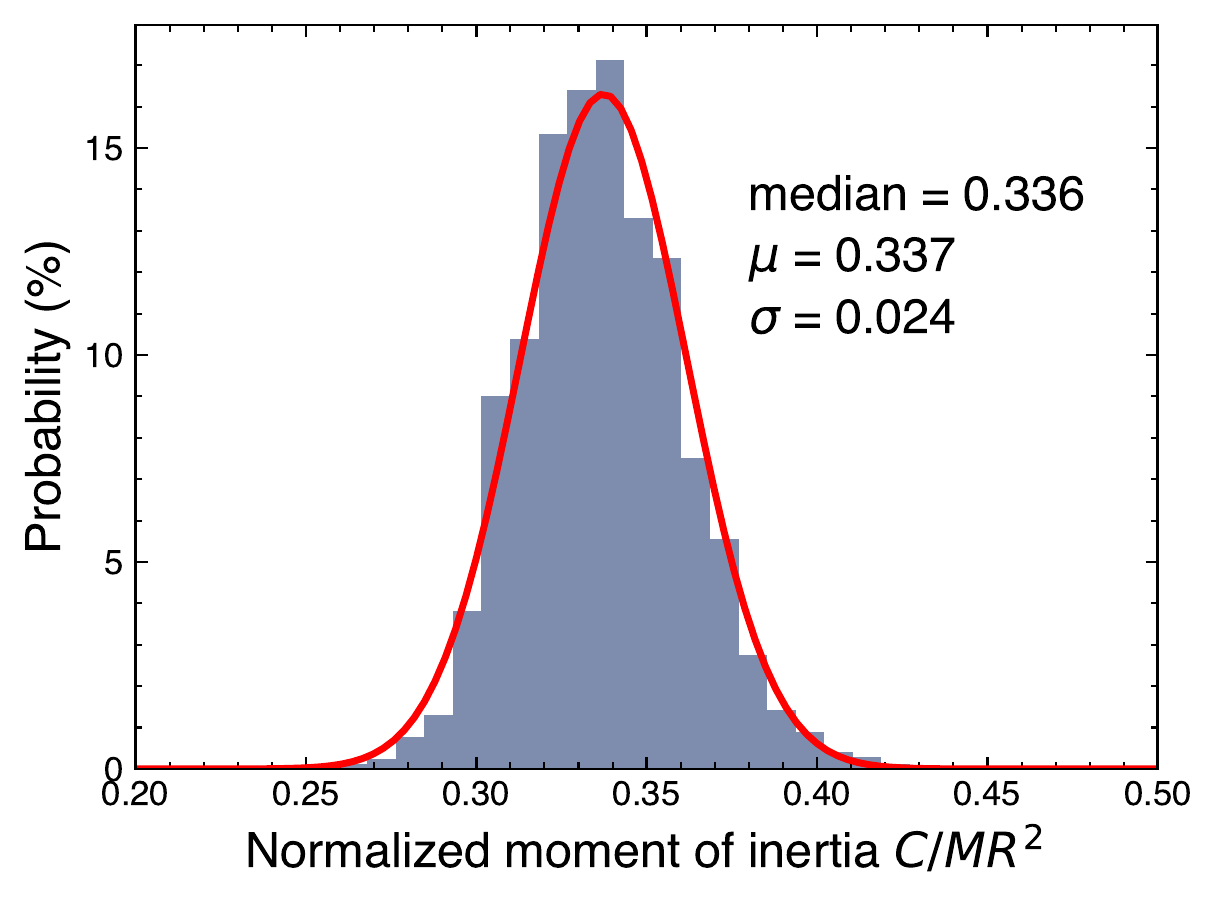} \\  
    \caption{
      A bootstrap analysis of radar speckle tracking estimates of the normalized moment of
      inertia of Venus suggests residual uncertainties of 7\% with the
      data obtained to date.}
  \label{fig-boot}
\end{figure}

The distribution of normalized moments of inertia from the bootstrap
analysis suggests residual uncertainties of 7\% with the data obtained
to date
(Fig.~\ref{fig-boot}).
The results are not yet sufficient to rule out
certain classes of interior models, whose normalized moments of inertia
computed in a recent study \cite{dumo16} span the range 0.327
-- 0.342.  Nevertheless, the best-fit value of the moment of inertia
factor combined with knowledge of the bulk density ($\rho$=5242.8
kg/m$^3$) enable a crude estimate of the size of the core of Venus with a
two-layer, uniform-density model
(See `Two-layer interior structure model' in Methods).
We find a core radius of approximately 3500 km (58\% of the planetary
radius) with large ($>$500 km) uncertainties due to both model
limitations and current uncertainties on $C/MR^2$.

\paragraph*{Spin period and length-of-day variations}

\begin{figure}[p]
  \centering
  \includegraphics[angle=0,width=3.1in]{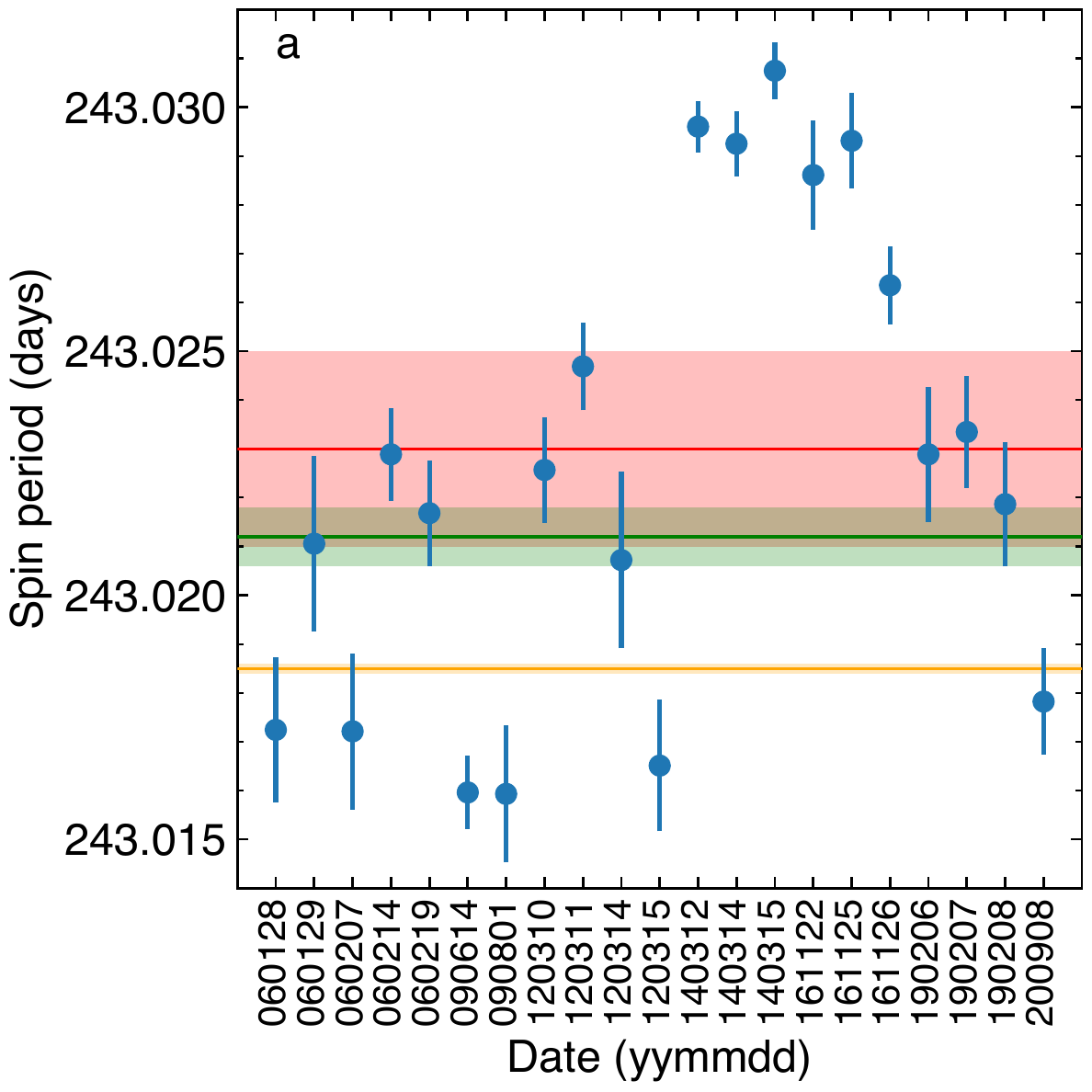} 
  \includegraphics[angle=0,width=3.1in]{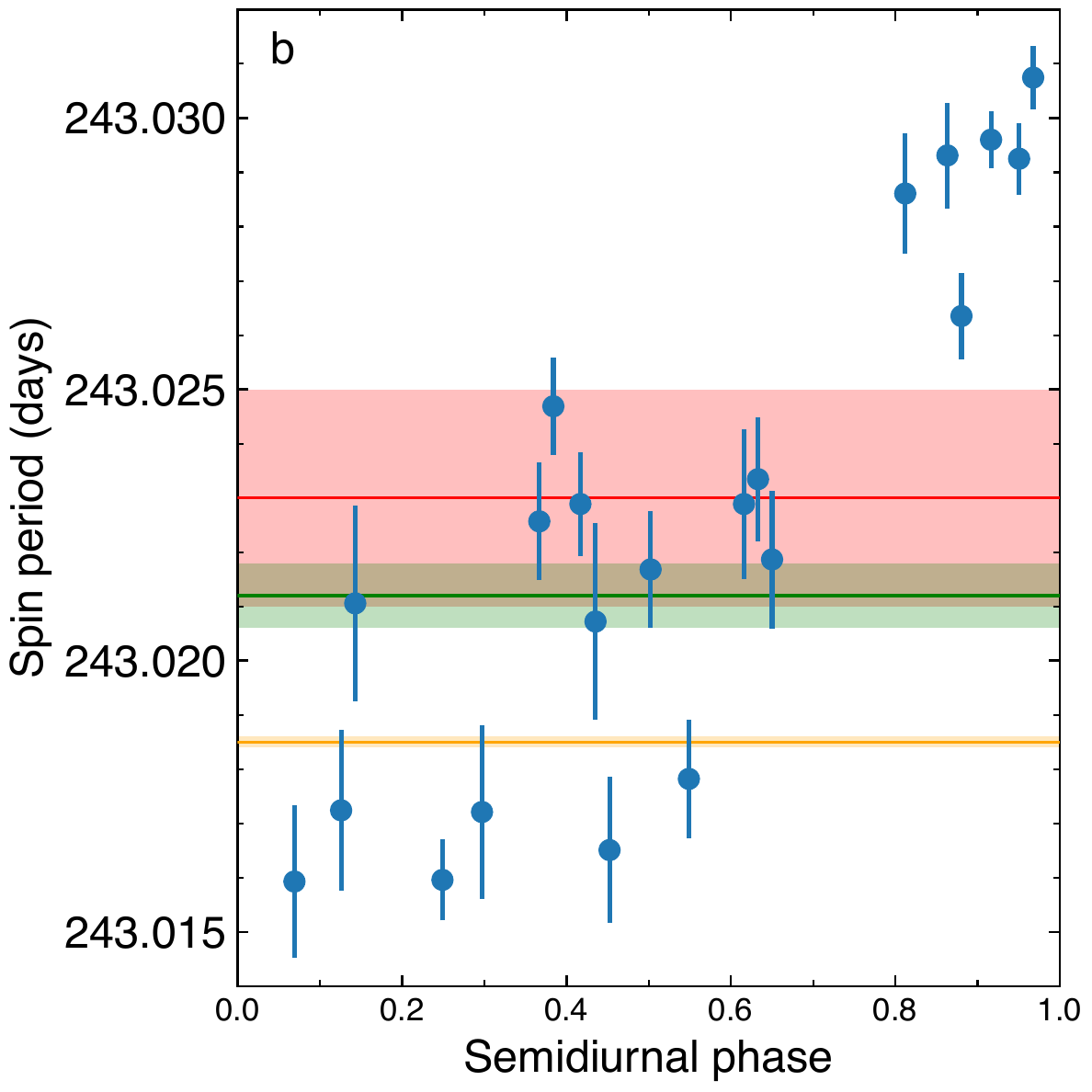} 
    \includegraphics[angle=0,width=3.1in]{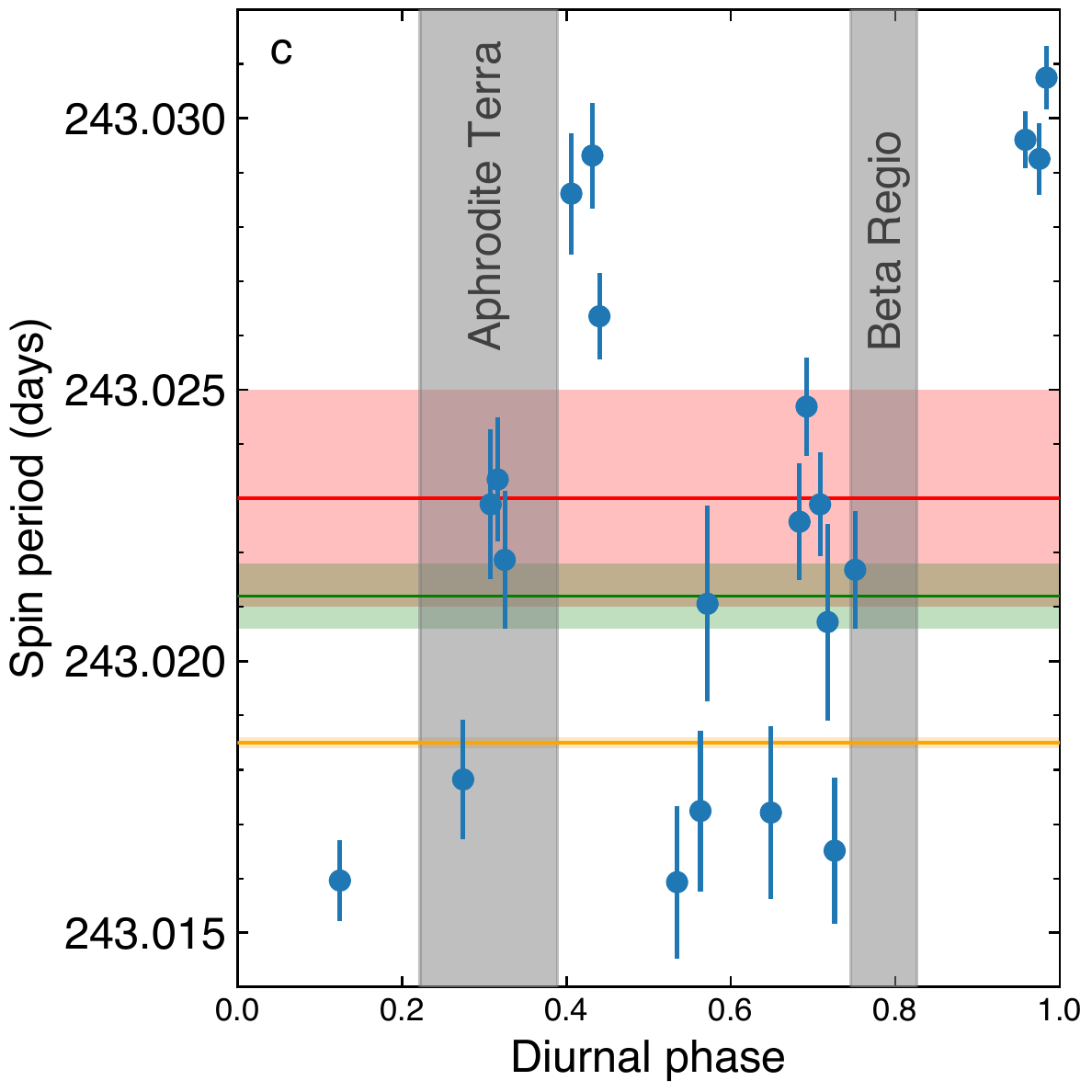}
  \includegraphics[angle=0,width=3.1in]{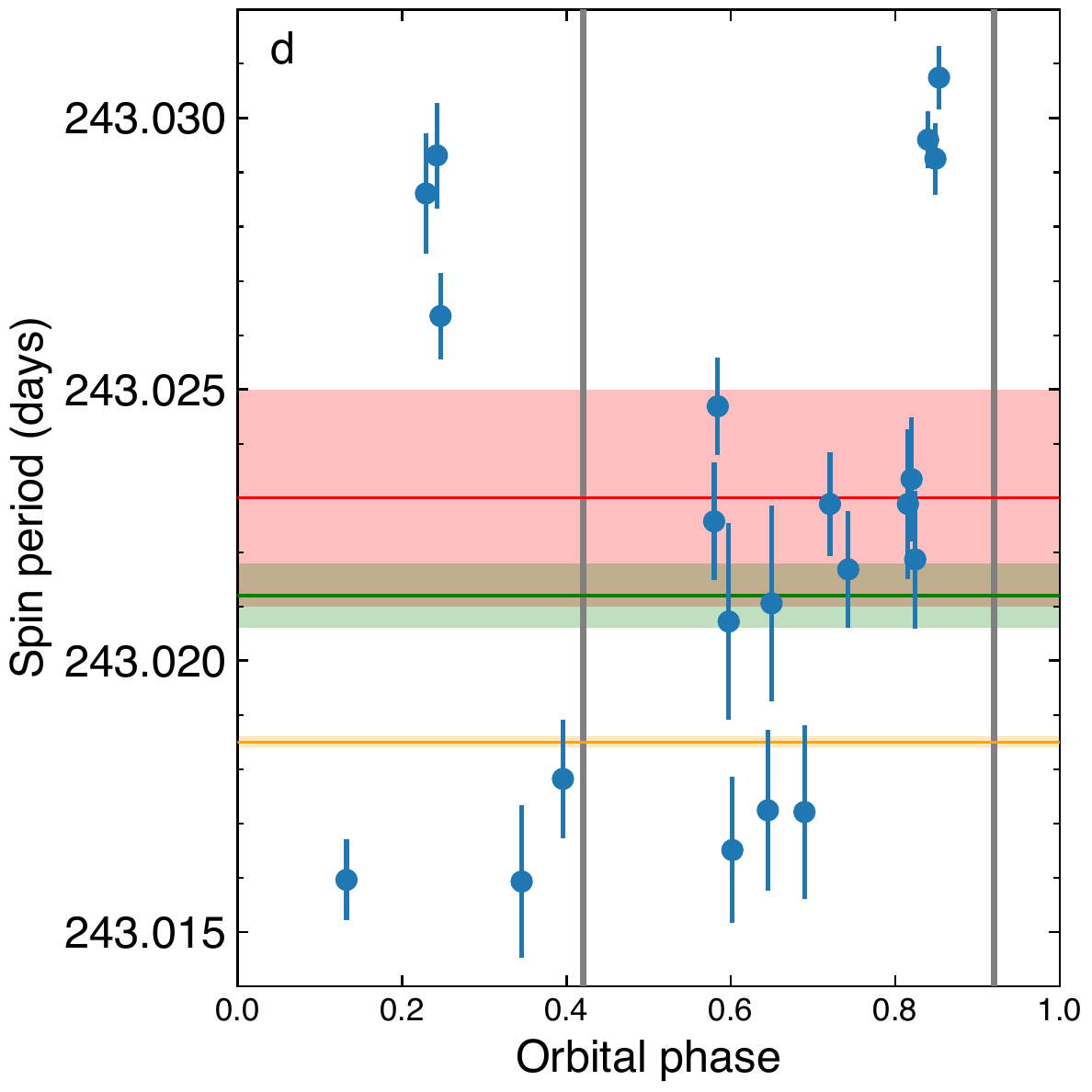}
    \caption{ ({\bf a}) Measurements of the {\em instantaneous} spin
      period of Venus shown as blue dots with $1\sigma$ error bars.
      These data are based on
      time lag measurements of 
      radar speckles observed with Goldstone and the GBT between 2006 and 2020.  Horizontal lines
      and shaded areas show the {\em average} spin periods derived
      from Magellan radar images (1990--1992, orange), Magellan and
      Venus Express images (1991 and 2007, red), and Earth-based radar
      images (1988--2017, green).  Same data folded onto a ({\bf b})
      semidiurnal cycle ($\sim$58.4 days), ({\bf c}) diurnal cycle
      ($\sim$116.8 days), and ({\bf d}) orbital cycle ($\sim$224.7
      days).
      The origin of phase is 2000 FEB 22
      10:43:58 TDB, when the sub-solar longitude is approximately
      zero.  Vertical grey bands and lines indicate local noon over low-latitude
      highlands and equinoxes, respectively.  }
    \label{fig-lod}
\end{figure}

We used the time lag measurements
to compute the spin period of Venus at each observation epoch
(Table \ref{tab-obs}).  The data show that Venus
exhibits substantial LOD variations (Fig.~\ref{fig-lod}).  We reject
the hypothesis of a constant spin period with high confidence
(p-value $<<10^{-16}$)
because a model with constant spin period yields a large sum of squares of residuals (SSR=620).
In addition, we find that published values
of Venus's average spin rate
\cite{davi92,muel12,camp19} are
inconsistent with most of our instantaneous spin period measurements.

\enlargethispage{0.5cm}

Our data set spans almost 15 years and includes 21 measurements with an
average fractional uncertainty of 5 ppm.  %
The median value of our measurements provides a robust estimate of the
{\em average} length of day on Venus, $P$=243.0226 $\pm$ 0.0013 days ($1\sigma$),
where the error bars are obtained by bootstrap resampling
(see `Estimate of average spin period' in Methods).
Natural variability around the mean is
at least $\pm$ 0.0047 days ($1\sigma$).
Our improved determinations of the spin axis orientation, precession
rate, and spin period form the basis of a recommended orientation model for
Venus (Supplementary Information).
This model and the model currently in use \cite{davi92} yield differences in the predicted inertial positions of equatorial landmarks that grow by $\sim 1$ km per year.
Stochastic LOD variations over a 10-year period contribute an
additional uncertainty of $\pm$3.3 km ($1\sigma$), which will
complicate the establishment of new geodetic control networks and the
measurement of the
spin precession from
orbital or landed platforms (Supplementary Information).

The fractional excursion in instantaneous spin rate observed to date
is 61 ppm, which corresponds to variations in spin period of 0.015
days or 21 minutes.
On seven instances, we observed Venus on consecutive
days and measured variations ranging between 2 and 17 ppm with a
weighted average value of 9 $\pm$ 5 ppm ($1\sigma$),
suggesting a spin rate of change as large as
$d\omega/dt \sim 3.1 \times 10^{-17}$~rad~s$^{-2}$    %
and corresponding torques
$T = C d\omega/dt \sim 1.9 \times 10^{21}$~N~m.
The
LOD variations observed at Venus are 3 orders of magnitude larger than
on Earth,
where core-mantle interactions can change the LOD by $\sim$4~ms
(46~ppb) on $\sim$20-year timescales \cite{gros07}. The torques
responsible for the LOD variations on Earth are $T_\oplus =C_\oplus d\omega_{\oplus}/dt
\sim 4.3 \times 10^{17}$~N~m, where $d\omega_{\oplus}/dt \sim 5.4
\times 10^{-21}$~rad~s$^{-2}$ and $C_\oplus = 8.0 \times
10^{37}$~kg~m$^2$.  If Venus has a liquid core, it may experience
torques of the same order of magnitude,
which would yield $d\omega/dt
\sim
10^{-20}$~rad~s$^{-2}$,
a factor of
$10^3$ too small
compared to observations.
Tidal despinning torques are an order of magnitude smaller.  We
conclude that changes in AAM are primarily responsible for the LOD
variations at Venus.
Other contributions to the LOD variations include a $\sim$3 ppm variation
at semidiurnal frequencies due to solar torques on Venus's permanent
deformation and possibly sub-ppm variations due to core-mantle interactions
\cite{cott11}.
The AAM variations are so large that they likely prevent capture in resonances with Earth,
a phenomenon that has been hypothesized for decades
\cite{gold66resonant,inge78,yode97,bill05venus}.

If the AAM budget on Venus is $2.9 \times 10^{28}$~kg~m$^2$~s$^{-1}$
\cite{sanc17}, the 61 ppm fractional excursion in spin period measured
to date provides a lower bound on the fractional change in AAM of
$\epsilon = 3.8\%$.  
consecutive Earth days correspond to spin period changes of $\sim$3
minutes and $\epsilon=0.6\%$.  
in AAM amounting to $\epsilon=0.1\%$ or less over half a Venusian
day~\cite{lebo10,cott11}, which suggests a rate of change in AAM that
is $\sim$300 times lower than what we observe.  A more recent GCM
simulation indicates $\epsilon$ as large as 0.6\% over a quarter of a
Venusian day~\cite{nava18}, or a rate of change that is $\sim$30 times
lower than what we observe.  Modeling the dynamical state of the Venus
atmosphere with quantitative precision is difficult~\cite{sanc17} and
validation of Venus GCMs is complicated by the fact that few
measurements of the internal dynamics of the atmosphere are
available. Our measurements provide useful calibration data for GCMs.

Because secular evolution of the spin
rate is likely, we tested for the presence of a linear trend in our
measurements.  The slopes detected in linear regressions with
unweighted and weighted uncertainties are not statistically
significant at $p=0.066$ and $p=0.078$, respectively, and we are
unable to detect a long-term trend in the data obtained to date.

We also examined whether the measurements exhibited periodicities
related to the spin ($\omega$), orbital ($n$), and diurnal ($\omega +
n$) frequencies, including
$\omega$, $n$, $2\omega$, $\omega + n$, $2n$, $\omega + 2n$,  $2\omega+ 2n$ (Fig.~\ref{fig-lod}).
Although we are currently unable to detect periodicities with
confidence or attribute the LOD variations to specific causes, we
speculate on possible causes and effects of periodicities at
semidiurnal, diurnal, or orbital periods that may be present
(Supplementary Fig.~\ref{fig-Lomb} and Supplementary Tables
\ref{tab-Lombphys} and \ref{tab-PDMphys}).

The possibility of semidiurnal variations in the AAM budget is intriguing
because this period controls the strength of the atmospheric tide
\cite{gold69venus,inge78,dobr80}.  If confirmed, this periodicity
would require an impulsive release of
$\sim$4\% of the total AAM twice per Venusian day, with profound
consequences for the internal dynamics of the atmosphere.  The
rotation of Venus would repeatedly slow down over a period of 58.4
days and depart from the equilibrium value while AAM increased.  An
impulsive release of AAM would provide a restoring torque and spin the
solid planet back up. How such a process would operate is unknown.

If diurnal variations in AAM were confirmed instead, one could invoke
mountain torques to explain some of the variations.  Mountain torques
are hypothesized to cause remarkable planetary-scale features observed
in the rapidly rotating upper atmosphere that are stationary with
respect to the slowly rotating surface \cite{fuku17,kouy17}.  The
torques affect the rotation rate of the solid body~\cite{nava18}, and
our data suggest slower rotation after midday over low-latitude
highlands.  However, the simulations conducted to date suggest diurnal
changes in LOD of 2 minutes~\cite{nava18}, whereas
we observe 
LOD variations of at least 20 minutes.

One possible cause of AAM variations on annual timescales is the
convective updraft related to Hadley cell circulation, whose location
depends on sub-solar latitude \cite{mitchell09}.  With thermal
inertia, the maxima in AAM and minima in spin period would be
delayed from the equinoxes (Fig.~\ref{fig-lod}).

\paragraph*{Methods}

\small

\paragraph*{Radar speckle tracking}

When Venus is illuminated with a monochromatic radio signal, a large
number of individual surface and near-surface elements scatter the
signal back towards the observer,
where each contribution to the radar echo has a specific amplitude and phase.
It is the superposition
(complex sum) of these individual responses that
gives the radar echo its speckled nature.  Because of constructive and
destructive interference, the echo power varies in a random-like
fashion (Supplementary Fig.~\ref{fig-speckles}).  However, apart from
receiver noise, the received signal is not random but is determined by
the distribution and properties of scatterers on the rigid surface of
Venus.  Therefore, the pattern of speckles is tied to the rotation of
Venus and sweeps over the surface of the Earth along a trajectory
dictated by the spin state.  Green \cite{gree62,gree68} described this
pattern as frozen corrugations in the reflected wavefront and
illustrated it by drawing contours of constant electric field strength
moving in the receiver (ground) plane.  He also detailed how the
motion of the pattern is diagnostic of the target's instantaneous spin
state and suggested cross-correlating time series of the electric
field amplitudes recorded at two receiving stations.
The speckle coherence conditions and applicability to the measurement
of planetary spin states were expanded by Holin \cite{holi88,holi92},
who also showed that the technique works for arbitrary
topography.

The characteristic scale of the speckles is given by the classic
diffraction formula $r\lambda/d$, where $r$ is the range to the
planet, $\lambda$ is the wavelength, and $d$ is the diameter of the
scattering area on the target, on the order of the planetary radius.
For Venus ($R$=6051.8~km) observed at $r$=0.8 astronomical units (au)
and $\lambda$=3.5~cm, the speckle scale is $\sim$0.7~km.  
Most of the time, observers located at separate antennas record
distinct radar speckle patterns, in which case
cross-correlation of the radar echo time series obtained at separate
antennas yields low correlation scores.  During brief periods
($\sim$30~s) on suitable days, the wavefront corrugations follow a
trajectory that sweeps over both antennas used in this work
(Supplementary Fig.~\ref{fig-path}), the Goldstone and Green Bank
antennas.  When this situation arises, large correlation scores
($>$0.6) are obtained at certain time lags, typically $\sim$20~s.
The {\em
  epoch} at which the high correlation occurs is diagnostic of the
spin axis orientation.  The {\em time lag} at which the high
correlation occurs yields a measurement of the instantaneous spin
period.   

The short duration of the high-correlation condition is explained by
the speckle size and the length of the projected baseline, i.e., the
antenna separation vector projected perpendicular to the line of
sight.  Because the speckle scale ($\sim$0.7~km) is so small compared
to the projected baseline ($\sim$3000 km), a small (0.01 deg)
misalignment of the speckle trajectory with respect to the baseline
orientation results in appreciable decorrelation.  For an east-west
baseline that oscillates daily by $\pm$23 deg with respect to the
ecliptic, the high correlation condition is maintained each day for
only $\sim$30 s.

Radar speckle tracking was used to reveal that Mercury's outer core is
molten \cite{marg07} and
to measure its moment of inertia and core size
\cite{marg12jgr,marg18merc}.  The
accuracy of the technique was
demonstrated by subsequent spacecraft measurements of Mercury's spin
axis orientation and amplitude of longitude librations, which
are in excellent agreement (1\%) with the radar estimates
\cite{star15grl}.

\paragraph*{Observing protocol}

We illuminated Venus with monochromatic radiation (8560 MHz, 450
kW) from the Deep Space Network (DSN) 70~m antenna in Goldstone,
California (DSS-14), and we recorded the speckle pattern as it swept
over two receiving stations (DSS-14 and the 100~m antenna in Green
Bank, West Virginia).  The transmitted waveform was circularly
polarized (right-circular, IEEE definition), and we recorded the
echoes in both right-circular (same sense, SC) and left-circular
polarizations (opposite sense, OC).  The OC echo is generally an order
of magnitude stronger and was used for the spin state measurements.
To compensate for the Earth-Venus Doppler shift, the transmitted
waveform was continuously adjusted in frequency by a programmable
local oscillator
so that the echo center at the Green Bank Telescope (GBT) remained
fixed at 8560 MHz.  Because the Doppler shift was compensated for
reception at the GBT, there was a residual Doppler shift during
reception at Goldstone.  Differential Doppler corrections were applied
with a programmable local oscillator at Goldstone so that the echo
center also remained fixed at 8560 MHz.  At both stations, a positive
frequency offset of 2000 Hz was added in the frequency downconversion
chain to prevent the $\sim$350-Hz-wide Venus echo from overlapping
with 0 Hz (DC).

On any given day, transmission typically occurred for the duration of
the round-trip light time to Venus.  The receive window started
immediately after transmission ended.  Transmit times were selected so
that the predicted high-correlation epochs were positioned within the
receive windows.

During reception, variations in the electric field were detected by
the standard low-noise X-band receivers at DSS-14 and GBT.
At Goldstone, the signal was converted to intermediate frequencies of
325 MHz and 50 MHz prior to mixing to baseband.  At GBT, 
the signal was converted to intermediate frequencies of 720 MHz and 30
MHz prior to mixing to baseband.  During conversion to baseband the
in-phase (I) and quadrature (Q) components of the signal were
generated.  Both I and Q voltages were low-pass filtered, sampled with
analog-to-digital converters, and recorded to computer hard drives.
For most observations, voltages were low-pass filtered at 1.9 MHz,
sampled at 5 MHz by our custom-built Portable Fast Sampler (PFS)
data-taking systems~\cite{pfs}, and stored with 4-bit resolution.
The 2020 observations and the 2019 observations at the GBT were
low-pass filtered at 1.25 MHz, sampled at 6.25 MHz by the Dual Channel
Agile Receiver (DCAR) data-taking system, and stored with 8-bit
resolution.

\paragraph*{Data reduction technique}
After the observations, we downsampled the data to effective sampling
rates $f_s$ between 30~Hz and 5000 Hz and computed the complex
cross-correlation of the Goldstone and GBT signals.

The I and Q samples can be thought of as the real and imaginary parts
of a complex signal $\left\{ z(t) \right\}$, with $z(t)=I(t)+jQ(t)$
and $j=\sqrt{-1}$.

The complex-valued cross-correlation of the signals $\left\{ z_1(t)
\right\}$ and $\left\{ z_2(t) \right\}$ is given by
\begin{equation}
R_{z_1z_2}(t,\tau) = E[z_1(t) z_2^*(t+\tau)],
\end{equation}
where E[] represents the expectation value operator, and * represents
the complex conjugate operator.  The normalized value of the
correlation is obtained with
\begin{equation}
\rho_{z_1z_2}(t,\tau) = \frac{|R_{z_1 z_2}(t,\tau)|}{\sqrt{|R_{z_1z_1}(t,0)||R_{z_2z_2}(t,0)|}},
\end{equation}
where $||$ is the absolute value operator.  Since the maximum possible
value of the correlation $R_{zz}(t,\tau)$ at each $t$ occurs at $\tau=0$,
$\rho_{z_1z_2}(t,\tau)$ is $\leq 1$ for all $\tau$.

The complex cross-correlation $\rho_{z_1z_2}(t,\tau)$ is a
two-dimensional correlation function in the variables $t$ (epoch) and
$\tau$ (time lag).  Examples of one-dimensional slices through the
peak of the correlation function are shown in
Fig.~\ref{fig-corr}.  We fit Gaussians to the one-dimensional slices
to obtain estimates of the epochs of correlation maximum $\hat t$.  We
also obtained estimates of the time lags $\hat\tau$ that maximize the
correlation functions.

For epoch correlations, we used $f_s$ = 200 Hz, about half the Doppler
broadening due to Venus's rotation, and integration times of 4 s,
except for the 2020 data where integration times of 1 s were used due
to reduced phase coherence related to the lack of Doppler compensation
on that day.
We also obtained correlations with $f_s$ = 200 Hz after low-pass
filtering of the radar echoes with a cutoff frequency set at 10\% of
the Doppler broadening.  We found that the low-pass filtered versions,
which have higher overall signal-to-noise ratios, yielded slightly
larger correlation values than the unfiltered versions, 0.65 compared
to 0.59 on average, and we used them in the analysis.

We assigned uniform uncertainties to epoch measurements for two
reasons.  First, the width of the Gaussian correlates with speckle
size and therefore Earth-Venus distance (Supplementary Information).  Because
certain baseline orientations are only observable at certain
distances, non-uniform uncertainties would bias the fit towards
certain baseline orientations.  Second, complex correlations of the
2012 observations are corrupted, but amplitude correlations are
well-behaved. We used the amplitude correlations results for the 2012
data, but amplitude correlations have notably narrower Gaussian widths
than complex correlations.

For time lag correlations, we used $f_s$ = 5000 Hz, boxcar averaging
to 200 Hz, and non-overlapping integration times of 1 s within +/- 10
s of the peak, which yielded 21 independent estimates during the high
correlation period.  We selected all estimates with a correlation
amplitude larger than 0.3, which left 12--21 points per epoch (19 data
points on average) and a mean correlation amplitude of 0.58.

The time lags evolve appreciably with time due to the changing
Earth-Venus geometry, and we performed linear regressions of the time
lag measurements to produce one estimate of the time lag
per epoch.  We used 2000 bootstrap trials to
randomly exclude data points from the linear regressions and used the
bootstrap means and standard deviations as estimates of the time lags
and uncertainties at the reference epochs (Table \ref{tab-obs}).  The
intrinsic variability of these estimates estimates is 2--7 ppm and 4
ppm on average.

Certain systematic effects may affect the spin rate measurements.  We
evaluated the error in spin period determination introduced by the
residual uncertainty on spin axis orientation by solving for spin
periods at various orientations within the error ellipse.  We found
that it is $<$1~ppm for 15 epochs and $<$2~ppm for the remaining 6
epochs.  We evaluated the error introduced by imperfect knowledge of
the epoch of correlation maximum by solving for spin periods with
epochs modified by the standard deviation of epoch residuals
($\pm$0.32~s) multiplied by $\sqrt{N/(N-M)}$ to adjust for the sample
standard deviation (here, $N=21$ and $M=3$).  We found that it is $<$3 ppm for
17 epochs and $<$5 ppm for the remaining 4 epochs, with the largest
errors affecting the March 2012 observations.  We added the variances
due to these errors to the variances due to bootstrap resampling of
the time lag measurements to produce spin period uncertainties.  The
resulting fractional uncertainties range from 2 ppm to 7 ppm with an
average value of 5 ppm.

\paragraph*{Conversion of observables to spin state estimates}

We used the observables $\hat t$ and $\hat \tau$ to obtain spin state
estimates (spin axis orientation and instantaneous spin rates).  In
these calculations the planet state vectors are furnished by a
Navigation and Ancillary Information Facility (NAIF) kernel (de438.bsp) that
represents the Jet Propulsion Laboratory Planetary Ephemeris
DE438.
The Earth orientation is provided by a NAIF kernel (earth\_latest\_high\_prec.bpc) that includes
up-to-date timing and polar motion
data.
The formalism for predicting the ($t$,$\tau$) values that yield high
correlations is described in detail in Appendix B of \cite{marg12jgr}.
Calculations include time delays accounting for light-travel times,
general relativistic corrections to the time delays, and Lorentz
transformations for bounce point conditions \cite{marg12jgr}.
We link the observables $\hat t$ and $\hat \tau$ to spin state
estimates with these predictions and the following procedures.

The space-time positions of the two receiving stations at the epochs
of correlation maxima were used to solve for the spin axis orientation
that generates similar speckles at both receiving stations.  We used a
least-squares approach to minimize the residuals between the predicted
epochs and the observed epochs.  The best-fit spin axis orientation is
referred to the epoch J2000.0.  The precession model for the spin axis
is given by equation (\ref{eq-precess}).

After the spin axis orientation was determined, we used each time lag
measurement to determine the instantaneous spin rate at the
corresponding epoch, once again based on the similarity requirement
for the speckles.  We iteratively adjusted the nominal spin rate of
-1.4813688 deg/day by a multiplicative factor until the predicted time
lag matched the observed time lag.

The nominal DSN-GBT baseline is 3260 km in length.  In the spin rate
problem, it is the projected baseline that is relevant, i.e., the
baseline component that is perpendicular to the line of sight.
Because of the displacement of the light rays due to refraction in the
Earth's atmosphere, the effective projected baseline differs from the nominal
value.  A correction factor for refraction within Earth's atmosphere
(Table~\ref{tab-geom}) was applied to the spin rate at each
observation epoch.  The formalism for this calculation is described in
detail in Appendix C of \cite{marg12jgr}.  However, these corrections
are small.  The worst case correction at the largest zenith angle of
$\sim 65^\circ$ is $\sim$15~m for a projected baseline of $\sim$3257
km, i.e., a fractional change of 4 ppm.

Passage through Venus's atmosphere retards light rays to both
telescopes by $<$1 microsecond, which is much smaller than our
uncertainties.  Our measurements are robust with respect to refraction
within Venus's atmosphere because the light rays to each scatterer
follow essentially identical paths during the high correlation epoch.
The differences in incidence angles for scatterers observed by DSN and
GBT always differ by $<$6 arcseconds and most differ by $<$2.5
arcseconds, such that the light rays received at DSN and GBT
experience essentially identical atmospheric delays.

One-way absorption through the atmosphere of Venus at the sub-Earth
point is $\sim$5.62~dB at X band \cite{duan10}, which effectively
decreases our signal-to-noise ratio by a factor of $\sim$10 compared
to that of a hypothetical atmosphereless Venus.

\paragraph*{Two-layer interior structure model}

We considered a two-layer, uniform-density model to provide a crude estimate of the core size.  We emphasize the limitations of such a model.  The large pressures inside Venus result in density profiles that vary with depth, which violate the uniform density assumptions.  The two-layer model therefore yields biased estimates.

The three unknowns are the bulk density of the core, the bulk density of the mantle, and the radius of the core.  Dumoulin et al.\ [2] used a rescaled version of the Preliminary Reference Earth Model (PREM) \cite{prem} to estimate Venus core densities.  Perhaps in part as a result of this choice, all of their models with $C/MR^2$ in the range 0.327 -- 0.342 have core densities within 1\% of 10358 kg/m$^3$.  We set the core density to this value and solved for the other two unknowns, being mindful that other assumptions on core density would yield different results.  We found a core radius of 3508 km (58\% of the planetary radius) and mantle density of 4006~kg/m$^3$.  However, current uncertainties on $C/MR^2$ result in large (500~km) uncertainties on core size.  For comparison, the Earth's core radius is 3480 km (55\% of Earth's equatorial radius) and mantle density is $\sim$4400~kg/m$^3$.

\paragraph*{Estimate of average spin period}

Our data set spans almost 15 years and includes 21 measurements of the
instantaneous spin period with an average fractional uncertainty of 5
ppm.  The median of our instantaneous measurements, $P = 243.0226$
days, provides a robust estimate of the average spin period, which we
confirmed with 10,000 bootstrap trials.  In these trials, mock data
sets were created by selecting 21 data points at random, with
replacement.  These trials demonstrate robustness with respect to
inclusion or exclusion of certain data points.  For each trial, we
computed the weighted average and the median of the spin period.  The
distribution of weighted averages yields 243.0227 $\pm$ 0.0013 days
(95\% C.I., 243.0202 -- 243.0252 days) and the distribution of medians
yields 243.0224 $\pm$ 0.0012 (95\% C.I., 243.0207 -- 243.0247 days),
with a median of 243.0226 days.  We adopt $243.0226 \pm 0.0013$ days
($1\sigma$) as our best estimate of the average spin period of Venus
in the interval 2006--2020.

Our estimate differs substantially from the $\sim$500-day-average
Magellan estimate of 243.0185 $\pm$ 0.0001 days~\cite{davi92}, is
almost identical to the $\sim$16-year-average spin period estimate of
243.023 $\pm$ 0.001 days of Mueller et al.~\cite{muel12}, and is
marginally consistent with the $\sim$29-year-average spin period
estimate of 243.0212 $\pm$ 0.0006 days of Campbell et
al.~\cite{camp19}.

\paragraph*{Identification of periodicities}
We tested for the presence of periodicities by computing Lomb
periodograms \cite{press92} and phase dispersion minima (PDM)
\cite{PDM}.  We tested thousands of trial periods between 1 day and
5400 days (Supplementary Fig.~\ref{fig-Lomb5400}).  These analyses
yielded ranked lists of candidate periods, the first ten of which were
examined and found to have no obvious physical significance
(Supplementary Tables \ref{tab-Lomb5400} and \ref{tab-PDM}).  Phase
folding the data with the candidate periods did not result in
convincing patterns (Supplementary Fig.~\ref{fig-top4}), suggesting
that these candidate periodicities are spurious detections from a
noisy and sparsely sampled data set.

We also examined whether the measurements exhibited periodicities
related to the spin ($\omega$), orbital ($n$), and diurnal ($\omega +
n$) frequencies, including
$\omega$, $n$, $2\omega$, $\omega + n$, $2n$, $\omega + 2n$,  $2\omega+ 2n$.
The semidiurnal period ranked highest according to both the Lomb
periodogram (Supplementary Fig.~\ref{fig-Lomb} and Supplementary Table
\ref{tab-Lombphys}) and the $\theta_{\rm PDM}$ statistic, which also
favored the diurnal period and, to a lesser extent,
the orbital period (Supplementary Table \ref{tab-PDMphys}).

\pagebreak
\bibliography{notes}

\begin{thebibliography}{10}
\expandafter\ifx\csname url\endcsname\relax
  \def\url#1{\texttt{#1}}\fi
\expandafter\ifx\csname urlprefix\endcsname\relax\def\urlprefix{URL }\fi
\providecommand{\bibinfo}[2]{#2}
\providecommand{\eprint}[2][]{\url{#2}}

\bibitem{smre18}
\bibinfo{author}{Smrekar, S.~E.}, \bibinfo{author}{Davaille, A.} \&
  \bibinfo{author}{Sotin, C.}
\newblock \href{http://dx.doi.org/10.1007/s11214-018-0518-1}{{Venus Interior
  Structure and Dynamics}}.
\newblock \emph{\bibinfo{journal}{Space Sci. Rev.}}
  \textbf{\bibinfo{volume}{214}}, \bibinfo{pages}{88} (\bibinfo{year}{2018}).

\bibitem{dumo16}
\bibinfo{author}{Dumoulin, C.}, \bibinfo{author}{Tobie, G.},
  \bibinfo{author}{Verhoeven, O.}, \bibinfo{author}{Rosenblatt, P.} \&
  \bibinfo{author}{Rambaux, N.}
\newblock \href{http://dx.doi.org/10.1002/2016JE005249}{{Tidal constraints on
  the interior of Venus}}.
\newblock \emph{\bibinfo{journal}{J. Geophys. Res. Planets}}
  \textbf{\bibinfo{volume}{122}}, \bibinfo{pages}{1338--1352}
  (\bibinfo{year}{2017}).

\bibitem{orou18}
\bibinfo{author}{{O'Rourke}, J.~G.}, \bibinfo{author}{{Gillmann}, C.} \&
  \bibinfo{author}{{Tackley}, P.}
\newblock \href{http://dx.doi.org/10.1016/j.epsl.2018.08.055}{{Prospects for an
  ancient dynamo and modern crustal remanent magnetism on Venus}}.
\newblock \emph{\bibinfo{journal}{Earth Planet. Sci. Lett.}}
  \textbf{\bibinfo{volume}{502}}, \bibinfo{pages}{46--56}
  (\bibinfo{year}{2018}).

\bibitem{davi92}
\bibinfo{author}{{Davies}, M.~E.} \emph{et~al.}
\newblock \href{http://dx.doi.org/10.1029/92JE01166}{{The rotation period,
  direction of the north pole, and geodetic control network of Venus}}.
\newblock \emph{\bibinfo{journal}{J. Geophys. Res.}}
  \textbf{\bibinfo{volume}{97}}, \bibinfo{pages}{13141--}
  (\bibinfo{year}{1992}).

\bibitem{muel12}
\bibinfo{author}{{Mueller}, N.~T.}, \bibinfo{author}{{Helbert}, J.},
  \bibinfo{author}{{Erard}, S.}, \bibinfo{author}{{Piccioni}, G.} \&
  \bibinfo{author}{{Drossart}, P.}
\newblock \href{http://dx.doi.org/10.1016/j.icarus.2011.09.026}{{Rotation
  period of Venus estimated from Venus Express VIRTIS images and Magellan
  altimetry}}.
\newblock \emph{\bibinfo{journal}{Icarus}} \textbf{\bibinfo{volume}{217}},
  \bibinfo{pages}{474--483} (\bibinfo{year}{2012}).

\bibitem{camp19}
\bibinfo{author}{{Campbell}, B.~A.} \emph{et~al.}
\newblock \href{http://dx.doi.org/10.1016/j.icarus.2019.06.019}{{The mean
  rotation rate of Venus from 29 years of Earth-based radar observations}}.
\newblock \emph{\bibinfo{journal}{Icarus}} \textbf{\bibinfo{volume}{332}},
  \bibinfo{pages}{19--23} (\bibinfo{year}{2019}).

\bibitem{yode97}
\bibinfo{author}{{Yoder}, C.~F.}
\newblock {Venusian Spin Dynamics}.
\newblock In \emph{\bibinfo{booktitle}{Venus II: Geology, Geophysics,
  Atmosphere, and Solar Wind Environment}}, \bibinfo{pages}{1087--}
  (\bibinfo{publisher}{U. of Az. Press}, \bibinfo{year}{1997}).

\bibitem{sanc17}
\bibinfo{author}{{S{\'a}nchez-Lavega}, A.}, \bibinfo{author}{{Lebonnois}, S.},
  \bibinfo{author}{{Imamura}, T.}, \bibinfo{author}{{Read}, P.} \&
  \bibinfo{author}{{Luz}, D.}
\newblock \href{http://dx.doi.org/10.1007/s11214-017-0389-x}{{The Atmospheric
  Dynamics of Venus}}.
\newblock \emph{\bibinfo{journal}{Space Sci. Rev.}}
  \textbf{\bibinfo{volume}{212}}, \bibinfo{pages}{1541--1616}
  (\bibinfo{year}{2017}).

\bibitem{hori20}
\bibinfo{author}{{Horinouchi}, T.} \emph{et~al.}
\newblock \href{http://dx.doi.org/10.1126/science.aaz4439}{{How waves and
  turbulence maintain the super-rotation of Venus{\textquoteright}
  atmosphere}}.
\newblock \emph{\bibinfo{journal}{Science}} \textbf{\bibinfo{volume}{368}},
  \bibinfo{pages}{405--409} (\bibinfo{year}{2020}).

\bibitem{vexag19}
\bibinfo{author}{{Venus Exploration Assessment Group}}.
\newblock {Roadmap for Venus Exploration} (\bibinfo{year}{2019}).

\bibitem{kaul68short}
\bibinfo{author}{Kaula, W.~M.}
\newblock \emph{\bibinfo{title}{{An Introduction to Planetary Physics: The
  Terrestrial Planets}}} (\bibinfo{publisher}{Wiley}, \bibinfo{year}{1968}).

\bibitem{will94}
\bibinfo{author}{{Williams}, J.~G.}
\newblock \href{http://dx.doi.org/10.1086/117108}{{Contributions to the Earth's
  obliquity rate, precession, and nutation}}.
\newblock \emph{\bibinfo{journal}{Astron. J.}} \textbf{\bibinfo{volume}{108}},
  \bibinfo{pages}{711--724} (\bibinfo{year}{1994}).

\bibitem{folk97}
\bibinfo{author}{{Folkner}, W.~M.}, \bibinfo{author}{Yoder, C.~F.},
  \bibinfo{author}{Yuan, D.~N.}, \bibinfo{author}{Standish, E.~M.} \&
  \bibinfo{author}{{Preston}, R.~A.}
\newblock \href{http://dx.doi.org/10.1126/science.278.5344.1749}{{Interior
  structure and seasonal mass redistribution of Mars from Radio Tracking of
  Mars Pathfinder}}.
\newblock \emph{\bibinfo{journal}{Science}} \textbf{\bibinfo{volume}{278}},
  \bibinfo{pages}{1749--1751} (\bibinfo{year}{1997}).

\bibitem{stan13}
\bibinfo{author}{Standish, E.~M.} \& \bibinfo{author}{Williams, J.~G.}
\newblock Orbital ephemerides of the sun, moon, and planets.
\newblock In \bibinfo{editor}{Urban, S.~E.} \& \bibinfo{editor}{Seidelmann,
  P.~K.} (eds.) \emph{\bibinfo{booktitle}{Explanatory Supplement to the
  Astronomical Almanac}} (\bibinfo{publisher}{University Science Books},
  \bibinfo{year}{2013}).

\bibitem{kono99}
\bibinfo{author}{{Konopliv}, A.~S.}, \bibinfo{author}{{Banerdt}, W.~B.} \&
  \bibinfo{author}{{Sjogren}, W.~L.}
\newblock \href{http://dx.doi.org/10.1006/icar.1999.6086}{{Venus Gravity: 180th
  Degree and Order Model}}.
\newblock \emph{\bibinfo{journal}{Icarus}} \textbf{\bibinfo{volume}{139}},
  \bibinfo{pages}{3--18} (\bibinfo{year}{1999}).

\bibitem{cott09}
\bibinfo{author}{{Cottereau}, L.} \& \bibinfo{author}{{Souchay}, J.}
\newblock \href{http://dx.doi.org/10.1051/0004-6361/200912174}{{Rotation of
  rigid Venus: a complete precession-nutation model}}.
\newblock \emph{\bibinfo{journal}{Astron.\ Astrophys.}}
  \textbf{\bibinfo{volume}{507}}, \bibinfo{pages}{1635--1648}
  (\bibinfo{year}{2009}).

\bibitem{saun92}
\bibinfo{author}{{Saunders}, R.~S.} \emph{et~al.}
\newblock \href{http://dx.doi.org/10.1029/92JE01397}{{Magellan mission
  summary}}.
\newblock \emph{\bibinfo{journal}{J. Geophys. Res.}}
  \textbf{\bibinfo{volume}{97}}, \bibinfo{pages}{13067--13090}
  (\bibinfo{year}{1992}).

\bibitem{shap79}
\bibinfo{author}{{Shapiro}, I.~I.}, \bibinfo{author}{{Campbell}, D.~B.} \&
  \bibinfo{author}{{de Campli}, W.~M.}
\newblock \href{http://dx.doi.org/10.1086/182975}{{Nonresonance rotation of
  Venus}}.
\newblock \emph{\bibinfo{journal}{Astrophys. J. Lett.}}
  \textbf{\bibinfo{volume}{230}}, \bibinfo{pages}{L123--L126}
  (\bibinfo{year}{1979}).

\bibitem{zoha80}
\bibinfo{author}{{Zohar}, S.}, \bibinfo{author}{{Goldstein}, R.~M.} \&
  \bibinfo{author}{{Rumsey}, H.~C.}
\newblock \href{http://dx.doi.org/10.1086/112775}{A new radar determination of
  the spin vector of {V}enus}.
\newblock \emph{\bibinfo{journal}{Astronomical Journal}}
  \textbf{\bibinfo{volume}{85}}, \bibinfo{pages}{1103--1111}
  (\bibinfo{year}{1980}).

\bibitem{shap90}
\bibinfo{author}{{Shapiro}, I.~I.}, \bibinfo{author}{{Chandler}, J.~F.},
  \bibinfo{author}{{Campbell}, D.~B.}, \bibinfo{author}{{Hine}, A.~A.} \&
  \bibinfo{author}{{Stacy}, N.~J.~S.}
\newblock \href{http://dx.doi.org/10.1086/115602}{{The spin vector of Venus}}.
\newblock \emph{\bibinfo{journal}{Astron. J.}} \textbf{\bibinfo{volume}{100}},
  \bibinfo{pages}{1363--1368} (\bibinfo{year}{1990}).

\bibitem{slad90}
\bibinfo{author}{{Slade}, M.~A.}, \bibinfo{author}{{Zohar}, S.} \&
  \bibinfo{author}{{Jurgens}, R.~F.}
\newblock \href{http://dx.doi.org/10.1086/115603}{{Venus - Improved spin vector
  from Goldstone radar observations}}.
\newblock \emph{\bibinfo{journal}{Astron. J.}} \textbf{\bibinfo{volume}{100}},
  \bibinfo{pages}{1369--1374} (\bibinfo{year}{1990}).

\bibitem{gold66venus}
\bibinfo{author}{{Goldreich}, P.} \& \bibinfo{author}{{Peale}, S.}
\newblock \href{http://dx.doi.org/10.1086/109947}{{Spin-orbit coupling in the
  solar system}}.
\newblock \emph{\bibinfo{journal}{Astron. J.}} \textbf{\bibinfo{volume}{71}},
  \bibinfo{pages}{425--} (\bibinfo{year}{1966}).

\bibitem{goldpeal67}
\bibinfo{author}{{Goldreich}, P.} \& \bibinfo{author}{{Peale}, S.}
\newblock \href{http://dx.doi.org/10.1086/110289}{{Spin-orbit coupling in the
  solar system. II. The resonant rotation of Venus}}.
\newblock \emph{\bibinfo{journal}{Astron. J.}} \textbf{\bibinfo{volume}{72}},
  \bibinfo{pages}{662--} (\bibinfo{year}{1967}).

\bibitem{gold69venus}
\bibinfo{author}{{Gold}, T.} \& \bibinfo{author}{{Soter}, S.}
\newblock \href{http://dx.doi.org/10.1016/0019-1035(69)90068-2}{{Atmospheric
  Tides and the Resonant Rotation of Venus}}.
\newblock \emph{\bibinfo{journal}{Icarus}} \textbf{\bibinfo{volume}{11}},
  \bibinfo{pages}{356--} (\bibinfo{year}{1969}).

\bibitem{inge78}
\bibinfo{author}{{Ingersoll}, A.~P.} \& \bibinfo{author}{{Dobrovolskis}, A.~R.}
\newblock \href{http://dx.doi.org/10.1038/275037a0}{{Venus' rotation and
  atmospheric tides}}.
\newblock \emph{\bibinfo{journal}{Nature}} \textbf{\bibinfo{volume}{275}},
  \bibinfo{pages}{37--} (\bibinfo{year}{1978}).

\bibitem{dobr80}
\bibinfo{author}{{Dobrovolskis}, A.~R.} \& \bibinfo{author}{{Ingersoll}, A.~P.}
\newblock \href{http://dx.doi.org/10.1016/0019-1035(80)90156-6}{{Atmospheric
  tides and the rotation of Venus. I - Tidal theory and the balance of
  torques}}.
\newblock \emph{\bibinfo{journal}{Icarus}} \textbf{\bibinfo{volume}{41}},
  \bibinfo{pages}{1--17} (\bibinfo{year}{1980}).

\bibitem{corr01}
\bibinfo{author}{{Correia}, A.~C.~M.} \& \bibinfo{author}{{Laskar}, J.}
\newblock \href{http://dx.doi.org/10.1038/35081000}{{The four final rotation
  states of Venus}}.
\newblock \emph{\bibinfo{journal}{Nature}} \textbf{\bibinfo{volume}{411}},
  \bibinfo{pages}{767--770} (\bibinfo{year}{2001}).

\bibitem{bill05venus}
\bibinfo{author}{{Bills}, B.~G.}
\newblock \href{http://dx.doi.org/10.1029/2003JE002190}{{Variations in the
  rotation rate of Venus due to orbital eccentricity modulation of solar tidal
  torques}}.
\newblock \emph{\bibinfo{journal}{J. Geophys. Res.}}
  \textbf{\bibinfo{volume}{110}}, \bibinfo{pages}{11007--}
  (\bibinfo{year}{2005}).

\bibitem{hide80}
\bibinfo{author}{{Hide}, R.}, \bibinfo{author}{{Birch}, N.~T.},
  \bibinfo{author}{{Morrison}, L.~V.}, \bibinfo{author}{{Shea}, D.~J.} \&
  \bibinfo{author}{{White}, A.~A.}
\newblock \href{http://dx.doi.org/10.1038/286114a0}{{Atmospheric Angular
  Momentum Fluctuations and Changes in the Length of the Day}}.
\newblock \emph{\bibinfo{journal}{Nature}} \textbf{\bibinfo{volume}{286}},
  \bibinfo{pages}{114--} (\bibinfo{year}{1980}).

\bibitem{lebo10}
\bibinfo{author}{{Lebonnois}, S.} \emph{et~al.}
\newblock \href{http://dx.doi.org/10.1029/2009JE003458}{{Superrotation of
  Venus' atmosphere analyzed with a full general circulation model}}.
\newblock \emph{\bibinfo{journal}{J. Geophys. Res. Planets}}
  \textbf{\bibinfo{volume}{115}}, \bibinfo{pages}{E06006}
  (\bibinfo{year}{2010}).

\bibitem{cott11}
\bibinfo{author}{{Cottereau}, L.}, \bibinfo{author}{{Rambaux}, N.},
  \bibinfo{author}{{Lebonnois}, S.} \& \bibinfo{author}{{Souchay}, J.}
\newblock \href{http://dx.doi.org/10.1051/0004-6361/201116606}{{The various
  contributions in Venus rotation rate and LOD}}.
\newblock \emph{\bibinfo{journal}{Astron.\ Astrophys.}}
  \textbf{\bibinfo{volume}{531}}, \bibinfo{pages}{A45} (\bibinfo{year}{2011}).

\bibitem{pari11}
\bibinfo{author}{{Parish}, H.~F.} \emph{et~al.}
\newblock \href{http://dx.doi.org/10.1016/j.icarus.2010.11.015}{{Decadal
  variations in a Venus general circulation model}}.
\newblock \emph{\bibinfo{journal}{Icarus}} \textbf{\bibinfo{volume}{212}},
  \bibinfo{pages}{42--65} (\bibinfo{year}{2011}).

\bibitem{pfs}
\bibinfo{author}{Margot, J.~L.}
\newblock \href{http://dx.doi.org/10.1142/S225117172150001X}{A data-taking
  system for planetary radar applications}.
\newblock \emph{\bibinfo{journal}{J. Astron. Instrum.}}
  \textbf{\bibinfo{volume}{10}} (\bibinfo{year}{2021}).

\bibitem{marg07}
\bibinfo{author}{{Margot}, J.~L.}, \bibinfo{author}{{Peale}, S.~J.},
  \bibinfo{author}{{Jurgens}, R.~F.}, \bibinfo{author}{{Slade}, M.~A.} \&
  \bibinfo{author}{{Holin}, I.~V.}
\newblock \href{http://dx.doi.org/10.1126/science.1140514}{{Large Longitude
  Libration of Mercury Reveals a Molten Core}}.
\newblock \emph{\bibinfo{journal}{Science}} \textbf{\bibinfo{volume}{316}},
  \bibinfo{pages}{710--714} (\bibinfo{year}{2007}).

\bibitem{marg12jgr}
\bibinfo{author}{{Margot}, J.~L.} \emph{et~al.}
\newblock \href{http://dx.doi.org/10.1029/2012JE004161}{{Mercury's moment of
  inertia from spin and gravity data}}.
\newblock \emph{\bibinfo{journal}{J. Geophys. Res. Planets}}
  \textbf{\bibinfo{volume}{117}} (\bibinfo{year}{2012}).

\bibitem{star15grl}
\bibinfo{author}{Stark, A.} \emph{et~al.}
\newblock \href{http://dx.doi.org/10.1002/2015GL065152}{{First MESSENGER
  orbital observations of Mercury's librations}}.
\newblock \emph{\bibinfo{journal}{Geophys. Res. Lett.}}
  \textbf{\bibinfo{volume}{42}}, \bibinfo{pages}{7881--7889}
  (\bibinfo{year}{2015}).

\bibitem{bend86}
\bibinfo{author}{Bendat, J.~S.} \& \bibinfo{author}{Piersol, A.~G.}
\newblock \emph{\bibinfo{title}{Random Data. Analysis and Measurement
  Procedures}} (\bibinfo{publisher}{Wiley}, \bibinfo{year}{1986}),
  \bibinfo{edition}{2nd, revised and expanded} edn.

\bibitem{corr03a}
\bibinfo{author}{{Correia}, A.~C.~M.}, \bibinfo{author}{{Laskar}, J.} \&
  \bibinfo{author}{{de Surgy}, O.~N.}
\newblock \href{http://dx.doi.org/10.1016/S0019-1035(03)00042-3}{{Long-term
  evolution of the spin of Venus I. theory}}.
\newblock \emph{\bibinfo{journal}{Icarus}} \textbf{\bibinfo{volume}{163}},
  \bibinfo{pages}{1--23} (\bibinfo{year}{2003}).

\bibitem{corr03b}
\bibinfo{author}{{Correia}, A.~C.~M.} \& \bibinfo{author}{{Laskar}, J.}
\newblock \href{http://dx.doi.org/10.1016/S0019-1035(03)00043-5}{{Long-term
  evolution of the spin of Venus II. numerical simulations}}.
\newblock \emph{\bibinfo{journal}{Icarus}} \textbf{\bibinfo{volume}{163}},
  \bibinfo{pages}{24--45} (\bibinfo{year}{2003}).

\bibitem{gros07}
\bibinfo{author}{Gross, R.}
\newblock
  \href{http://dx.doi.org/https://doi.org/10.1016/B978-044452748-6.00057-2}{{Earth
  Rotation Variations – Long Period}}.
\newblock In \bibinfo{editor}{Schubert, G.} (ed.)
  \emph{\bibinfo{booktitle}{Treatise on Geophysics}}, \bibinfo{pages}{239 --
  294} (\bibinfo{publisher}{Elsevier}, \bibinfo{address}{Amsterdam},
  \bibinfo{year}{2007}).

\bibitem{gold66resonant}
\bibinfo{author}{{Goldreich}, P.} \& \bibinfo{author}{{Peale}, S.~J.}
\newblock \href{http://dx.doi.org/10.1038/2091117a0}{{Resonant Rotation for
  Venus?}}
\newblock \emph{\bibinfo{journal}{Nature}} \textbf{\bibinfo{volume}{209}},
  \bibinfo{pages}{1117--1118} (\bibinfo{year}{1966}).

\bibitem{nava18}
\bibinfo{author}{Navarro, T.}, \bibinfo{author}{Schubert, G.} \&
  \bibinfo{author}{Lebonnois, S.}
\newblock \href{http://dx.doi.org/10.1038/s41561-018-0157-x}{{Atmospheric
  mountain wave generation on Venus and its influence on the solid planet's
  rotation rate}}.
\newblock \emph{\bibinfo{journal}{Nat. Geosci.}} \textbf{\bibinfo{volume}{11}},
  \bibinfo{pages}{487--491} (\bibinfo{year}{2018}).

\bibitem{fuku17}
\bibinfo{author}{Fukuhara, T.} \emph{et~al.}
\newblock \href{http://dx.doi.org/10.1038/ngeo2873}{{Large stationary gravity
  wave in the atmosphere of Venus}}.
\newblock \emph{\bibinfo{journal}{Nat. Geosci.}} \textbf{\bibinfo{volume}{10}},
  \bibinfo{pages}{85--88} (\bibinfo{year}{2017}).

\bibitem{kouy17}
\bibinfo{author}{{Kouyama}, T.} \emph{et~al.}
\newblock \href{http://dx.doi.org/10.1002/2017GL075792}{{Topographical and
  Local Time Dependence of Large Stationary Gravity Waves Observed at the Cloud
  Top of Venus}}.
\newblock \emph{\bibinfo{journal}{Geophys. Res. Lett.}}
  \textbf{\bibinfo{volume}{44}}, \bibinfo{pages}{12,098--12,105}
  (\bibinfo{year}{2017}).

\bibitem{mitchell09}
\bibinfo{author}{{Mitchell}, J.~L.}
\newblock \href{http://dx.doi.org/10.1088/0004-637X/692/1/168}{{Coupling
  Convectively Driven Atmospheric Circulation to Surface Rotation: Evidence for
  Active Methane Weather in the Observed Spin Rate Drift of Titan}}.
\newblock \emph{\bibinfo{journal}{Astrophys. J.}}
  \textbf{\bibinfo{volume}{692}}, \bibinfo{pages}{168--173}
  (\bibinfo{year}{2009}).

\bibitem{gree62}
\bibinfo{author}{Green, P.~E.}
\newblock {Radar Astronomy Measurement Techniques}.
\newblock \bibinfo{type}{Tech. Rep.} \bibinfo{number}{282},
  \bibinfo{institution}{MIT Lincoln Laboratory} (\bibinfo{year}{1962}).

\bibitem{gree68}
\bibinfo{author}{Green, P.~E.}
\newblock Radar astronomy.
\newblock chap. \bibinfo{chapter}{Radar Measurements}
  (\bibinfo{publisher}{McGraw-Hill}, \bibinfo{year}{1968}).

\bibitem{holi88}
\bibinfo{author}{Kholin~(Holin), I.~V.}
\newblock \href{http://dx.doi.org/10.1007/BF01043597}{Spatial-temporal
  coherence of a signal diffusely scattered by an arbitrarily moving surface
  for the case of monochromatic illumination}.
\newblock \emph{\bibinfo{journal}{Radiophys. Quant. Elec.}}
  \textbf{\bibinfo{volume}{31}}, \bibinfo{pages}{371--374}
  (\bibinfo{year}{1988}).
\newblock \bibinfo{note}{Translated from I. V. Holin, {\em Izvestiya Vysshikh
  Uchebnykh Zavedenii, Radiofizika}, {\bf 31}, pp. 515-518.}

\bibitem{holi92}
\bibinfo{author}{Kholin~(Holin), I.~V.}
\newblock \href{http://dx.doi.org/10.1007/BF01038312}{Accuracy of
  body-rotation-parameter measurement with monochromatic illumination and
  two-element reception}.
\newblock \emph{\bibinfo{journal}{Radiophys. Quant. Elec.}}
  \textbf{\bibinfo{volume}{35}}, \bibinfo{pages}{284--287}
  (\bibinfo{year}{1992}).
\newblock \bibinfo{note}{Translated from I. V. Holin, {\em Izvestiya Vysshikh
  Uchebnykh Zavedenii, Radiofizika}, {\bf 35}, pp. 433-439.}

\bibitem{marg18merc}
\bibinfo{author}{Margot, J.~L.}, \bibinfo{author}{Hauck, S.~A.},
  \bibinfo{author}{Mazarico, E.}, \bibinfo{author}{Padovan, S.} \&
  \bibinfo{author}{Peale, S.~J.}
\newblock \href{http://dx.doi.org/10.1017/9781316650684.005}{{Mercury's
  Internal Structure}}.
\newblock In \bibinfo{editor}{Solomon, S.~C.}, \bibinfo{editor}{Nittler, L.~R.}
  \& \bibinfo{editor}{Anderson, B.~J.} (eds.)
  \emph{\bibinfo{booktitle}{Mercury: The View after MESSENGER}},
  \bibinfo{pages}{85--113} (\bibinfo{publisher}{Cambridge University Press},
  \bibinfo{year}{2018}).

\bibitem{duan10}
\bibinfo{author}{Duan, X.}, \bibinfo{author}{Moghaddam, M.},
  \bibinfo{author}{Wenkert, D.}, \bibinfo{author}{Jordan, R.~L.} \&
  \bibinfo{author}{Smrekar, S.~E.}
\newblock \href{http://dx.doi.org/https://doi.org/10.1029/2009RS004169}{X band
  model of venus atmosphere permittivity}.
\newblock \emph{\bibinfo{journal}{Radio Science}} \textbf{\bibinfo{volume}{45}}
  (\bibinfo{year}{2010}).

\bibitem{prem}
\bibinfo{author}{{Dziewonski}, A.~M.} \& \bibinfo{author}{{Anderson}, D.~L.}
\newblock \href{http://dx.doi.org/10.1016/0031-9201(81)90046-7}{Preliminary
  reference earth model}.
\newblock \emph{\bibinfo{journal}{Phys. Earth Planet. Inter.}}
  \textbf{\bibinfo{volume}{25}}, \bibinfo{pages}{297--356}
  (\bibinfo{year}{1981}).

\bibitem{press92}
\bibinfo{author}{Press, W.~H.}, \bibinfo{author}{Teukolsky, S.~A.},
  \bibinfo{author}{Vetterling, W.~T.} \& \bibinfo{author}{Flannery, B.~P.}
\newblock \emph{\bibinfo{title}{Numerical Recipes in C}}
  (\bibinfo{publisher}{Cambridge University Press},
  \bibinfo{address}{Cambridge, USA}, \bibinfo{year}{1992}),
  \bibinfo{edition}{2nd} edn.

\bibitem{PDM}
\bibinfo{author}{{Williams}, P. K.~G.}, \bibinfo{author}{{Clavel}, M.},
  \bibinfo{author}{{Newton}, E.} \& \bibinfo{author}{{Ryzhkov}, D.}
\newblock \href{http://arxiv.org/abs/1704.001}{{pwkit: Astronomical utilities
  in Python}} (\bibinfo{year}{2017}).
\newblock \eprint{1704.001}.

\bibitem{arch18}
\bibinfo{author}{{Archinal}, B.~A.} \emph{et~al.}
\newblock \href{http://dx.doi.org/10.1007/s10569-017-9805-5}{{Report of the IAU
  Working Group on Cartographic Coordinates and Rotational Elements: 2015}}.
\newblock \emph{\bibinfo{journal}{Celestial Mechanics and Dynamical Astronomy}}
  \textbf{\bibinfo{volume}{130}}, \bibinfo{pages}{22} (\bibinfo{year}{2018}).

\bibitem{kell14}
\bibinfo{author}{{Kelly}, B.~C.}, \bibinfo{author}{{Becker}, A.~C.},
  \bibinfo{author}{{Sobolewska}, M.}, \bibinfo{author}{{Siemiginowska}, A.} \&
  \bibinfo{author}{{Uttley}, P.}
\newblock \href{http://dx.doi.org/10.1088/0004-637X/788/1/33}{{Flexible and
  Scalable Methods for Quantifying Stochastic Variability in the Era of Massive
  Time-domain Astronomical Data Sets}}.
\newblock \emph{\bibinfo{journal}{Astrophys. J.}}
  \textbf{\bibinfo{volume}{788}}, \bibinfo{pages}{33} (\bibinfo{year}{2014}).

\end{thebibliography}

\bibliographystyle{jlm_naturemag} %

\paragraph*{Acknowledgments}

This article is dedicated to the memory of Raymond F. Jurgens, who was
instrumental in acquiring the data for this work.
We thank
M. A. Slade, J. T. Lazio, T. Minter, K. O'Neil, and F. J. Lockman for
assistance with scheduling the observations.
We thank B.~A. Archinal, P.~M. Davis, S. Lebonnois, J.~L. Mitchell, and C.~F. Wilson for useful comments and A. Lam for assistance with Figure 1.
The Green Bank Observatory is a
facility of the National Science Foundation operated under cooperative
agreement by Associated Universities, Inc.  Part of this work was
supported by the Jet Propulsion Laboratory, operated by Caltech under
contract with NASA.
We are grateful for NASA's Navigation and Ancillary Information
Facility software and data kernels, which greatly facilitated this
research.
JLM was funded in part by NASA grants
NNG05GG18G, NNX09AQ69G, NNX12AG34G, 80NSSC19K0870.

\paragraph*{Author contributions}

JLM conducted the investigation and wrote the
software and manuscript.  DBC contributed to the methodology.
JDG, JSJ, LGS, FDG, and AB contributed to data acquisition.  All
authors reviewed and edited the manuscript.

\paragraph*{Competing interests}

The authors declare no competing interests.

  \renewcommand{\figurename}{Supplementary Figure}
  \renewcommand{\tablename}{Supplementary Table}
\begin{center}

\pagebreak  
  {\Large \bf Supplementary Information}

\Large{Spin state and moment of inertia of Venus}\\

      \vspace{0.5cm}
\large
\author
    {Jean-Luc Margot\footnote[1]{email: jlm@epss.ucla.edu}\,, Donald B. Campbell,
      Jon D. Giorgini,\\ Joseph S. Jao, Lawrence G. Snedeker, Frank D. Ghigo, Amber Bonsall \\

      \vspace{0.5cm}
}

\end{center}
\setcounter{figure}{0} \renewcommand{\thefigure}{\arabic{figure}}
\setcounter{table}{0} \renewcommand{\thetable}{\arabic{table}}
\setcounter{page}{0}

\pagebreak

\normalsize  

\section{Log of attempted measurements}

Supplementary Table~\ref{tab-all} documents the history of attempted measurements of the spin
state of Venus with radar speckle tracking.

\LTcapwidth=\textwidth
\begin{longtable}[]{@{}llll@{}}
  \caption{Log of attempted measurements.}\\
  \toprule
Number & Date & Status & Reason\tabularnewline
\midrule
\endhead
01 & 2002 OCT 18 & no data & DSS-14 time request denied (Mars
Odyssey)\tabularnewline
02 & 2002 OCT 19 & no data & DSS-14 time request denied (Mars
Odyssey)\tabularnewline
03 & 2002 OCT 20 & no data & DSS-14 time request denied (Mars
Odyssey)\tabularnewline
04 & 2002 OCT 21 & no data & DSS-14 time request denied
(maintenance)\tabularnewline
05 & 2002 OCT 22 & no data & DSS-14 time request denied (Mars
Odyssey)\tabularnewline
06 & 2002 OCT 23 & no data & DSS-14 time request denied (Voyager
1)\tabularnewline
07 & 2002 OCT 24 & no data & DSS-14 time request denied
(Ulysses)\tabularnewline
08 & 2002 OCT 25 & no data & DSS-14 time request denied (asteroid 1997
XF11)\tabularnewline
09 & 2002 OCT 26 & no data & DSS-14 time request denied (asteroid 1997
XF11)\tabularnewline
10 & 2002 OCT 27 & no data & DSS-14 time request denied (asteroid 1997
XF11)\tabularnewline
11 & 2002 OCT 28 & no data & DSS-14 time request denied
(maintenance)\tabularnewline
12 & 2002 OCT 29 & no data & DSS-14 time request denied (Voyager
1)\tabularnewline
13 & 2002 OCT 30 & no data & DSS-14 time request denied (Mars
Odyssey)\tabularnewline
14 & 2002 OCT 31 & no data & DSS-14 time request denied
(Ulysses)\tabularnewline
15 & 2002 NOV 01 & no data & DSS-14 time request denied (Mars
Odyssey)\tabularnewline
16 & 2002 NOV 02 & no data & DSS-14 time request denied
(Ulysses)\tabularnewline
17 & 2002 NOV 03 & no data & DSS-14 time request denied
(Ulysses)\tabularnewline
18 & 2002 NOV 04 & no data & DSS-14 time request denied
(maintenance)\tabularnewline
19 & 2002 NOV 05 & no data & DSS-14 time request denied (Mars
Odyssey)\tabularnewline
20 & 2002 NOV 06 & no data & DSS-14 time request denied (Mars
Odyssey)\tabularnewline
21 & 2002 NOV 07 & no data & DSS-14 time request denied
(maintenance)\tabularnewline
22 & 2002 NOV 08 & no data & DSS-14 time request denied (Mars
Odyssey)\tabularnewline
23 & 2002 NOV 09 & no data & DSS-14 time request denied
(GBRA)\tabularnewline
24 & 2002 NOV 10 & no data & DSS-14 time request denied
(GBRA)\tabularnewline
25 & 2002 NOV 11 & no data & DSS-14 time request denied
(maintenance)\tabularnewline
26 & 2002 NOV 12 & no data & DSS-14 time request denied
(Galileo/Ulysses)\tabularnewline
27 & 2002 NOV 13 & no data & DSS-14 time request denied
(Ulysses)\tabularnewline
28 & 2002 NOV 14 & no data & DSS-14 time request denied
(Ulysses)\tabularnewline
29 & 2002 NOV 15 & no data & DSS-14 time request denied
(Ulysses)\tabularnewline
30 & 2002 NOV 16 & no data & DSS-14 time request denied (Mars
Odyssey/GBRA)\tabularnewline
31 & 2002 NOV 17 & no data & DSS-14 time request denied
(GBRA)\tabularnewline
32 & 2002 NOV 18 & no data & DSS-14 time request denied
(maintenance)\tabularnewline
33 & 2003 JAN 05 & no data & Venus correlation too close to Mercury
transmission\tabularnewline
34 & 2004 MAR 29 & no data & DSS-14 time request denied (Mars
Exploration Rover)\tabularnewline
35 & 2004 MAR 30 & no data & DSS-14 time request denied (Mars
Exploration Rover)\tabularnewline
36 & 2004 MAR 31 & incomplete data & Recording started too late after Mercury transmission
\tabularnewline
37 & 2004 APR 01 & no data & Goldstone motor-generator set
defective\tabularnewline
38 & 2004 APR 02 & no data & Goldstone motor-generator set
removed\tabularnewline
39 & 2004 MAY 03 & no data & DSS-14 time request denied
(maintenance)\tabularnewline
40 & 2004 MAY 04 & no data & Goldstone motor-generator set in
repair\tabularnewline
41 & 2004 MAY 05 & no data & Goldstone motor-generator set in
repair\tabularnewline
42 & 2004 MAY 06 & no data & DSS-14 time request denied
(maintenance)\tabularnewline
43 & 2004 MAY 07 & no data & Goldstone motor-generator set in
repair\tabularnewline
44 & 2006 JAN 23 & no data & Goldstone filament and pointing
problems\tabularnewline
45 & 2006 JAN 28 & detection &\tabularnewline
46 & 2006 JAN 29 & detection &\tabularnewline
47 & 2006 FEB 07 & detection &\tabularnewline
48 & 2006 FEB 09 & no data & Goldstone mirror did not
switch to receive position\tabularnewline
49 & 2006 FEB 12 & no data & GBT buried under 9 inches of
snow\tabularnewline
50 & 2006 FEB 14 & detection &\tabularnewline
51 & 2006 FEB 19 & detection &\tabularnewline
52 & 2006 FEB 23 & no data & GBT azimuth track repair\tabularnewline
53 & 2006 FEB 25 & no data & GBT azimuth track repair\tabularnewline
54 & 2008 DEC 01 & no data & Goldstone air corridor monitoring software bug\tabularnewline
55 & 2008 DEC 02 & no data & DSS-14 radiation clearance
denied\tabularnewline
56 & 2008 DEC 03 & no data & DSS-14 radiation clearance
denied\tabularnewline
57 & 2008 DEC 04 & no data & DSS-14 time request denied
(MESSENGER)\tabularnewline
58 & 2009 MAY 19 & no data & DSS-14 time request denied
(Spitzer)\tabularnewline
59 & 2009 MAY 20 & no data & DSS-14 time request denied
(Spitzer)\tabularnewline
60 & 2009 MAY 21 & no data & DSS-14 time request denied
(training/testing exercise)\tabularnewline
61 & 2009 JUN 09 & corrupted data & Corrupted local oscillator at
Goldstone\tabularnewline
62 & 2009 JUN 13 & no data & Goldstone operator ended receive cycle
prematurely\tabularnewline
63 & 2009 JUN 14 & detection &\tabularnewline
64 & 2009 JUN 21 & corrupted data & Echo shifted by -7600 Hz at
GBT\tabularnewline
65 & 2009 AUG 01 & detection &\tabularnewline
66 & 2011 JAN 04 & no data & DSS-14 transmitter heat exchanger in
repair\tabularnewline
67 & 2011 JAN 05 & no data & DSS-14 transmitter heat exchanger in
repair\tabularnewline
68 & 2011 JAN 06 & no data & DSS-14 time request denied (Voyager
1)\tabularnewline
69 & 2011 JAN 07 & no data & DSS-14 time request denied (Voyager
1)\tabularnewline
70 & 2011 JAN 08 & no data & GBT not pointing due to defective
stow pin switch\tabularnewline
71 & 2011 JAN 09 & no data & DSS-14 make-up time request denied (New
Horizons)\tabularnewline
72 & 2011 JAN 10 & no data & DSS-14 make-up time request denied
(maintenance)\tabularnewline
73 & 2011 JAN 11 & no data & DSS-14 make-up time request denied (New
Horizons)\tabularnewline
74 & 2011 JAN 12 & no data & DSS-14 make-up time request denied
(Cassini)\tabularnewline
75 & 2011 JAN 13 & no data & DSS-14 make-up time request denied
(maintenance)\tabularnewline
76 & 2011 JAN 14 & no data & DSS-14 make-up time request denied
(Stardust)\tabularnewline
77 &
2012
MAR 10 & detection & \tabularnewline
78 &
2012
MAR 11 & detection & \tabularnewline
79 &
2012
MAR 12 & no data & X-band receiver frozen at GBT\tabularnewline
80 &
2012
MAR 13 & no data & X-band receiver frozen at GBT\tabularnewline
81 &
2012
MAR 14 & detection & Cmplx correl.\ anomalous, amplitude
correl.\ okay\tabularnewline
82 &
2012
MAR 15 & detection & Cmplx correl.\ anomalous, amplitude
correl.\ okay\tabularnewline
83 &
2012
MAR 16 & incomplete data & Data gap prevents reliable
estimate of correlation peak\tabularnewline
84 & 2012 MAR 16 & no data & DSS-14 time request denied
(maintenance)\tabularnewline
85 &
2014
MAR 12 & detection &\tabularnewline
86 &
2014
MAR 13 & no data & DSS-14 unable to transmit (klystron \#2 vac-ion
pblm)\tabularnewline
87 &
2014
MAR 14 & detection &\tabularnewline
88 &
2014
MAR 15 & detection &\tabularnewline
89 &
2014
MAR 16 & no data & DSS-14 unable to transmit (klystron \#2 vac-ion
pblm)\tabularnewline
90 &
2016
NOV 21 & corrupted data & Differential Doppler not working at
DSS-14\tabularnewline
91 &
2016
NOV 22 & detection &\tabularnewline
92 & 2016 NOV 23 & no data & DSS-14 time request denied
(maintenance)\tabularnewline
93 &
2016
NOV 25 & detection &\tabularnewline
94 &
2016
NOV 26 & detection &\tabularnewline
95 &
2019
FEB 04 & no data & Voltage not recorded due to defective EDT cable\tabularnewline
96 &
2019
FEB 05 & no data & Voltage not recorded due to defective EDT cable \tabularnewline
97 &
2019
FEB 06 & detection & Recording with DCAR at GBT\tabularnewline
98 &
2019
FEB 07 & detection & Recording with DCAR at GBT\tabularnewline
99 &
2019
FEB 08 & detection & Recording with DCAR at GBT\tabularnewline
100 & 2020 FEB 20 & no data & DSS-14 klystron defective\tabularnewline
101 & 2020 FEB 21 & no data & DSS-14 klystron defective\tabularnewline
102 & 2020 FEB 22 & no data & DSS-14 klystron defective\tabularnewline
103 & 2020 FEB 23 & no data & DSS-14 klystron defective\tabularnewline
104 & 2020 FEB 24 & no data & DSS-14 klystron defective\tabularnewline
105 & 2020 MAY 07 & no data & Goldstone-VLA. DSS-14 klystron defective
\tabularnewline
106 & 2020 MAY 08 & no data & Goldstone-VLA. DSS-14 klystron defective
\tabularnewline
107 & 2020 MAY 28 & no data & Goldstone-VLA. DSS-14 klystron defective
\tabularnewline
108 & 2020 MAY 29 & no data & Goldstone-VLA. DSS-14 klystron defective
\tabularnewline
109 & 2020 JUN 29 & no data & Goldstone-VLA. DSS-14 klystron defective
\tabularnewline
110 & 2020 JUN 30 & no data & Goldstone-VLA. DSS-14 klystron defective
\tabularnewline
111 & 2020 JUL 11 & no data & Goldstone-VLA. DSS-14 klystron defective
\tabularnewline
112 & 2020 JUL 12 & no data & Goldstone-VLA. DSS-14 klystron defective
\tabularnewline
113 & 2020 JUL 13 & no data & Goldstone-VLA. DSS-14 klystron defective
\tabularnewline
114 & 2020 JUL 14 & no data & Goldstone-VLA. DSS-14 klystron defective
\tabularnewline
115 & 2020 JUN 15 & no data & Goldstone-VLA. DSS-14 klystron defective
\tabularnewline
116 & 2020 AUG 09 & no data & Goldstone-VLA. DSS-14 klystron defective
\tabularnewline
117 & 2020 AUG 22 & no data & DSS-14 klystron defective\tabularnewline
118 & 2020 AUG 23 & no data & DSS-14 klystron defective\tabularnewline
119 & 2020 AUG 30 & no data & DSS-14 klystron defective\tabularnewline
120 &
2020
SEP 05 & corrupted data & Strong, time-variable RFI at DSN\tabularnewline
121 &
2020
SEP 08 & detection & No Doppler
compensation on transmit or receive\tabularnewline
& & &\tabularnewline
\bottomrule
\label{tab-all}
\end{longtable}

\pagebreak
\section{Observing circumstances}

We observed Venus successfully on 21 instances between 2006 and 2020,
with observing circumstances reported in Supplementary
Table~\ref{tab-geom}.  The full list of observing attempts is shown in
Supplementary Table~\ref{tab-all}.

\begin{table}[H]
\begin{tabular}{rrrrrrrr}
{\#} & 
{date} &       
{RTT} &   
{$B_{\rm proj}$} &
{$B_{\rm long}$} &
{el$_{\rm GBT}$} &
{el$_{\rm DSN}$} &
{refract} \\ 
& 
{\small yymmdd}&       
{(s)} &   
{(km)} &
{$(^\circ)$} &
{$(^\circ)$} &
{$(^\circ)$} &
{($\mu$)} \\
\hline
01  &  060128  &   299.3  &  3250.5  &  -163.5  & 35.5  &  32.6 &	-3.71   \\ 
02  &  060129  &   303.5  &  3256.8  &  -163.8  & 35.2  &  33.5 &	-4.48   \\ 
03  &  060207  &   349.0  &  3242.1  &  -164.1  & 32.7  &  36.3 &	 3.48   \\ 
04  &  060214  &   391.9  &  3248.3  &  -162.1  & 32.8  &  35.6 &	 2.35   \\ 
05  &  060219  &   425.3  &  3258.0  &  -159.8  & 33.4  &  34.4 &	 2.04   \\ 
06  &  090614  &   770.8  &  2358.5  &   -51.8  & 29.4  &  57.6 &	 3.25   \\ 
07  &  090801  &  1132.9  &  3219.3  &     0.5  & 72.2  &  60.7 &	 0.52   \\ 
08  &  120310  &   838.7  &  2361.7  &   -55.1  & 30.5  &  59.2 &	 3.01   \\ 
09  &  120311  &   831.3  &  2403.0  &   -54.0  & 31.5  &  60.2 &	 2.75   \\ 
10  &  120314  &   808.9  &  2520.0  &   -50.8  & 34.6  &  63.1 &	 2.14   \\ 
11  &  120315  &   801.4  &  2556.8  &   -49.7  & 35.6  &  64.0 &	 1.98   \\ 
12  &  140312  &   597.5  &  3100.0  &  -144.5  & 35.5  &  25.0 &	 2.51   \\ 
13  &  140314  &   612.9  &  3060.0  &  -142.7  & 35.5  &  23.8 &	 2.75   \\ 
14  &  140315  &   620.6  &  3038.4  &  -141.8  & 35.5  &  23.3 &	 2.88   \\ 
15  &  161122  &  1045.1  &  3258.7  &  -167.5  & 25.0  &  25.7 &	 2.25   \\ 
16  &  161125  &  1024.8  &  3257.3  &  -164.0  & 26.0  &  24.8 &	 2.26   \\ 
17  &  161126  &  1017.9  &  3254.3  &  -162.8  & 26.3  &  24.4 &	 2.28   \\ 
18  &  190206  &   908.9  &  3154.3  &  -177.2  & 24.5  &  32.9 &	 2.16   \\ 
19  &  190207  &   916.2  &  3171.1  &  -176.0  & 25.0  &  32.7 &	 2.08   \\ 
20  &  190208  &   923.4  &  3186.4  &  -174.9  & 25.4  &  32.5 &	 2.01   \\ 
21  &  200908  &   909.4  &  2421.0  &    32.1  & 58.6  &  30.7 &	 2.75   \\
    \end{tabular}                                                                                                            
    \caption{Observational circumstances.  Each row gives the session
      number, the UT date of observation, the round-trip light time
      (RTT) in seconds, the length of the projected baseline in km, the
      ecliptic longitude of the projected baseline in degrees, the elevation
      angles at GBT and DSN during reception, and the correction
      factor ($\alpha$) required to account for refraction in the
      Earth's atmosphere [expressed as $\mu=(\alpha-1) \times 10^6$].  }
    \label{tab-geom}
\end{table}

\pagebreak
\section{Correlation epoch residuals}
\begin{figure}[H]
  \centering
    \includegraphics[angle=0,width=4.5in]{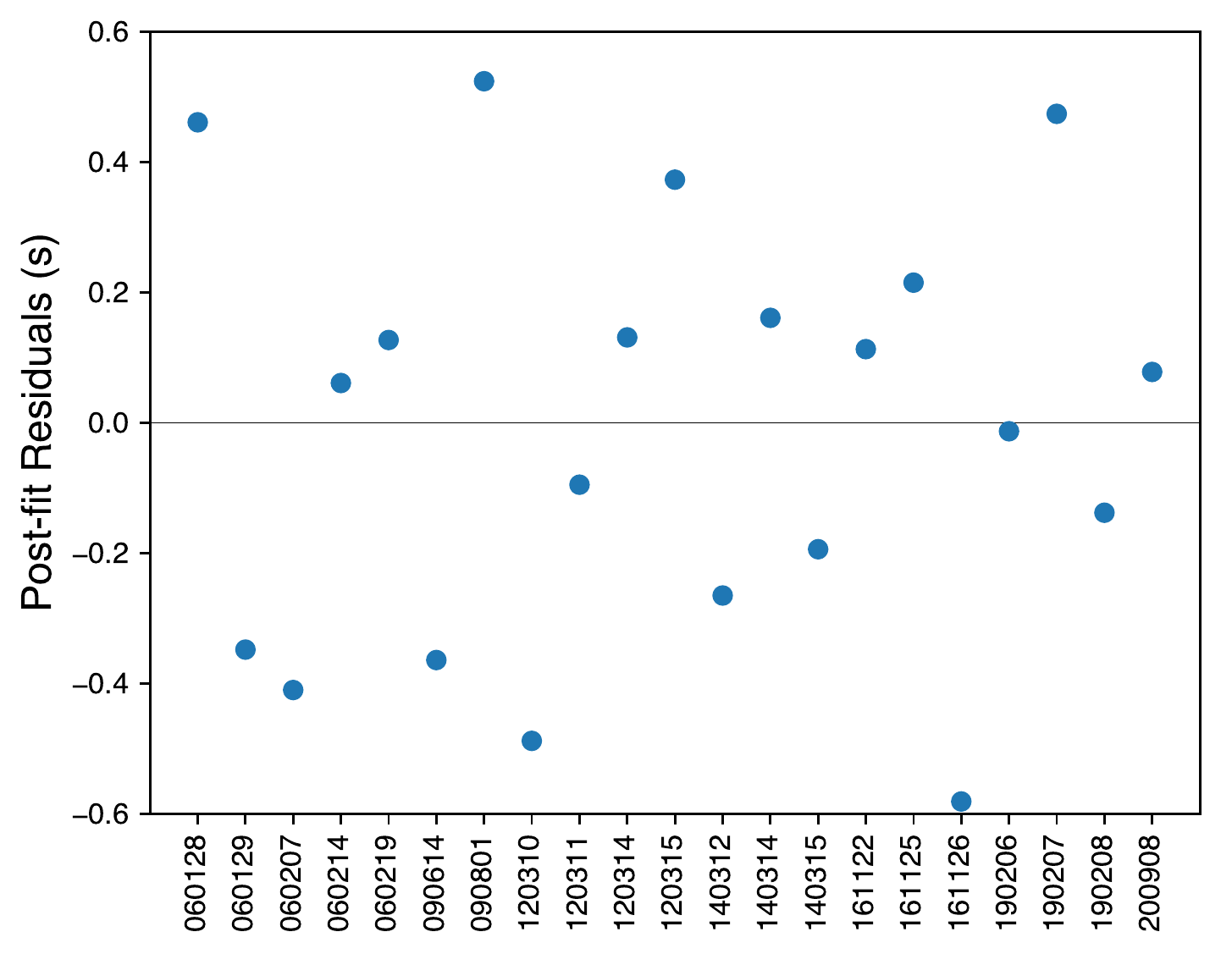} \\  
    \caption{Post-fit residuals, in seconds, from the spin axis
      orientation fit.  The standard deviation of residuals is 0.3 s.}
  \label{fig-residuals}
\end{figure}

Post-fit residuals have a standard deviation of 0.32 s and their
distribution is unremarkable (Supplementary Fig.~\ref{fig-residuals}).

\pagebreak
\section{Recommended orientation model for Venus}
\label{sec-iau}

The International Astronomical Union (IAU) provides orientation models
for planets and satellites.  In IAU-recommended models, the RA and DEC
of the spin axis in the International Celestial Reference Frame (ICRF)
and the longitude of the prime meridian (PM)\footnote{The RA and DEC represent the pole of
rotation that lies on the north side of the invariable plane of the
Solar System.  The longitude of the prime meridian is measured
easterly along the body's equator from the node at ${\rm RA}_0 + 90$
deg.} are represented in units
of degrees as quadratic polynomials of time, $x = x_o + x_1 t +
x_2 t^2$.  In the case of Venus, all coefficients of second order are
zero, and we can write
\begin{equation}
{\rm RA}  =  {\rm RA}_0  +  {\rm RA}_1 \times t/T,
\end{equation}
\begin{equation}
{\rm DEC} =  {\rm DEC}_0 + {\rm DEC}_1 \times t/T,
\end{equation}
\begin{equation}
  W   =  W_0   + W_1 \times t/d,
\end{equation}
where $t$ represents the number of seconds past the J2000.0 reference
epoch, $T$ represents the number of seconds per Julian century,
and $d$ represents the number of seconds per day.

The current IAU-recommended model for Venus \cite{arch18} is based on the Magellan
radar results \cite{davi92} and does not include precession rates for
the spin axis (RA$_1 = $ DEC$_1 = 0$):

\begin{verbatim}
BODY299_POLE_RA          = (  272.76       0.          0. )
BODY299_POLE_DEC         = (   67.16       0.          0. )
BODY299_PM               = (  160.20      -1.4813688   0. )
\end{verbatim}

Our recommended model is
\begin{verbatim}
BODY299_POLE_RA          = (  272.7391    -0.08154         0. )
BODY299_POLE_DEC         = (   67.1510    -0.04739         0. )
BODY299_PM               = (  160.28      -1.4813437       0. )
\end{verbatim}

The spin axis orientation of the recommended model differs from that
of the current IAU model by 0.012 deg.  As a result, predictions for
the inertial positions of features on the surface of Venus are off by
up to 1.3~km with the current IAU model.

In computing $W_1$, we used the robust estimate of $P$=243.0226 $\pm$
0.0013 days (this work), whose nominal value is in good agreement with
the $\sim$16-year-average spin period estimate of 243.023 $\pm$ 0.001
days~\cite{muel12}, but differs at the 6 ppm level from the
$\sim$29-year-average spin period estimate of 243.0212 $\pm$ 0.0006
days \cite{camp19}.  Because the estimates differ, it is reasonable to
inquire which estimate is most appropriate for $W_1$.
The 29-year-data set of Campbell et al.~\cite{camp19} includes 6
measurements with uncertainties of 0.028 deg and 0.042 deg at the
beginning and end of the 10,510-day interval, respectively.  In the
absence of systematic errors, the fractional error on the
29-year-average spin period could be as low as 3.2 ppm.  However,
ephemeris calculations used in the analysis relied on the current IAU
orientation, which biased the predicted locations of the tiepoints by
up to 1.3~km, more than twice the 600 m uncertainty that was assigned
to the tiepoint locations.  These uncertainties may also have been
underestimated for some of the tiepoint measurements because the
nominal radar range resolution of 600~m translates into a ground
resolution of $\sim$1 km for observations at incidence angles below 40
deg.  Finally, the tiepoints were observed at a variety of incidence
angles, with differences that occasionally exceeded 12 deg.  Given the
speckled nature of radar echoes and their strong dependence on
incidence angle, it may be difficult to locate the same tiepoint with
an accuracy of one resolution cell for echoes with such dissimilar
incidence angles and different noise realizations.  These
considerations may or may not affect the $\sim$29-year-average spin
period estimate in a substantial manner.  However, we are
unable to fully assess the possible impact of these potential biases
and uncertainties.
For this reason, we chose to use our estimate of the average spin period,
having demonstrated and quantified its robustness.  The
spin period value of $P=243.0226$ days yields
$W_1=-1.4813437$ deg/day, which differs from the current IAU
$W_1$
value by
$2.51 \times 10^{-5}$ deg/day.

In computing a revised value of $W_0$, we had to decide whether to use
``the most reliable estimate'' \cite{davi92} of the period, $P=
243.0185$ days, which corresponds to $W_1=-1.4813687$~deg/day, or
$W_1=-1.4813688$~deg/day, which is used in the current IAU orientation
model.  We chose the latter, which was reported by Merton Davies and
propagated in every subsequent IAU report, although the absence of
guard digits in the publication of reference \cite{davi92} prevents us
from knowing which value is correct.  We computed the orientation of
the prime meridian in degrees as of Jan 1, 1991 (Julian date
2448258.0) on the basis of the orientation at epoch J2000.0 (Julian
date 2451545.0) and the IAU-recommended $W_1$, as follows:
\begin{equation}
160.20 -1.4813688 \times (2448258.0 - 2451545.0) = 349.46
\end{equation}
We then propagated this orientation from Jan 1, 1991 to Jan 1, 2000,
with our recommended spin period, as follows:
\begin{equation}
349.46 +1.4813437  \times (2448258.0 - 2451545.0) = 160.28
\end{equation}
The difference in $W_0$ values of 0.083 deg at J2000.0 corresponds
to 8.8 km on the surface at the equator.  It is a plausible estimate
of the rotational error in the position of the prime meridian
accumulated between the Magellan epoch and J2000.0 if the period 
averages out to $P$=243.0226 days.  An additional error of
$2.51 \times 10^{-5}$ deg/day ($\sim$1 km/year) past J2000.0 must be
considered.  For instance, by Jan. 1, 2021,
the rotational error will have grown by an additional 0.192 deg to
a total of 0.275 deg, or 29.0 km at the equator.

Because of uncertainties in the spin period, the actual $W_0$ value
could be anywhere between 160.25 deg and 160.31 deg if we assume a
constant rate of rotation within the 1-sigma interval $P$=243.0226
$\pm$ 0.0013 days.
This uncertainty translates into a
$\pm 2.75$ km-wide region at the equator by
J2000.0.

\pagebreak
\section{Challenges associated with determining the inertial positions of landmarks and measuring the spin precession rate from orbital or landed platforms}

We quantified three challenges associated with determining the
inertial positions of specific landmarks on Venus.  These challenges
may complicate the targeting of landed missions, the measurement of
the spin precession rate from orbital or landed platforms, the
establishment of new geodetic control networks, and the
co-registration of remote sensing data sets.

First, predictions of the inertial positions of Venus landmarks with
an incorrect spin axis orientation result in a time-independent error
of up to 1.3~km.  Second, predictions with the average value of the
spin period currently recommended by the IAU results in an error that
increases with time at a rate of approximately $\sim$1 km/year.
Third, predictions with an incomplete knowledge of the spin period
history result in an uncertainty region that grows with time and can
reach $\pm3.3$ km ($1\sigma$) for 10-year-ahead forecasts.  The first
two problems can be mitigated to a large extent by adopting an updated
orientation model.  The third problem is more difficult to solve and
may place severe constraints on
the ability to convert between Venus-body-fixed and inertial
coordinates with high precision.

Our best-fit spin axis orientation differs from the IAU-recommended
spin axis orientation by 0.012 deg.  The absolute position of a
specific feature on the surface of Venus
in inertial space would be off by up to 1.3~km with the current IAU model.  Because the
obliquity of Venus is small and because the spin precession rate is
known with 7\% precision, this error will remain essentially constant
for decades.  In addition, this error can be almost entirely
eliminated by using a better estimate of the spin axis orientation.
The residual uncertainty in our spin axis orientation is $<$3 arcsec
and amounts to a maximum error of 80~m on the surface.

Likewise, our adopted spin rate differs from the IAU-recommended spin
rate by 0.0092 deg/year, which amounts to discrepancies in absolute
positions at the equator that accumulate at a rate of $\sim$1 km/year.
If the average spin period computed with data spanning 2006-2020 is
also representative of the average spin period since the Magellan
epoch ($\sim$1991), then this error can also be largely mitigated by
adopting our recommended orientation model (Section~\ref{sec-iau}).
The residual uncertainty in our average spin period value is
$\pm0.0013$ days ($1\sigma$)
and corresponds to errors at the equator that grow at a rate of
$\pm0.3$ km/year, assuming a constant spin period.

Unfortunately, LOD variations give rise to additional uncertainties in
determining the
inertial positions of surface features.  This
prediction error grows with time
and cannot be easily mitigated unless LOD variations are monitored
routinely.  We quantified the likely magnitude of this error by
fitting a continuous-time autoregressive moving average (CARMA) model
to our data and predicting the time evolution of the rotational phase
uncertainty.
CARMA models were developed to describe continuous-time processes
sampled at discrete times, including processes that include standard
Brownian motion.  They are appropriate for the analysis of sparsely
and irregularly sampled noisy time series that represent stationary
physical processes, such as our observations of the Venus LOD.

\begin{figure}[p]
  \centering
  \includegraphics[angle=0,width=6in]{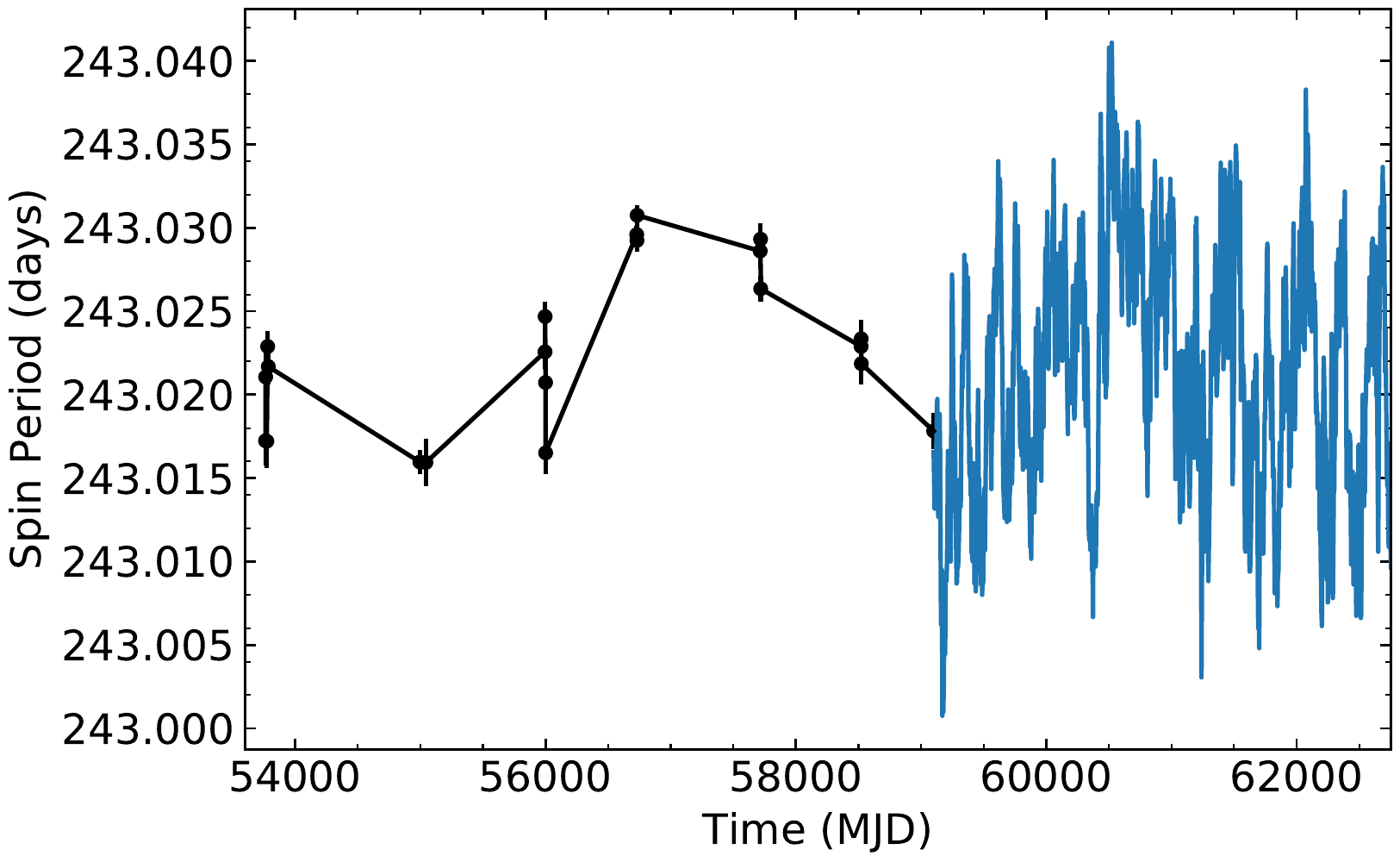}
  \includegraphics[angle=0,width=6in]{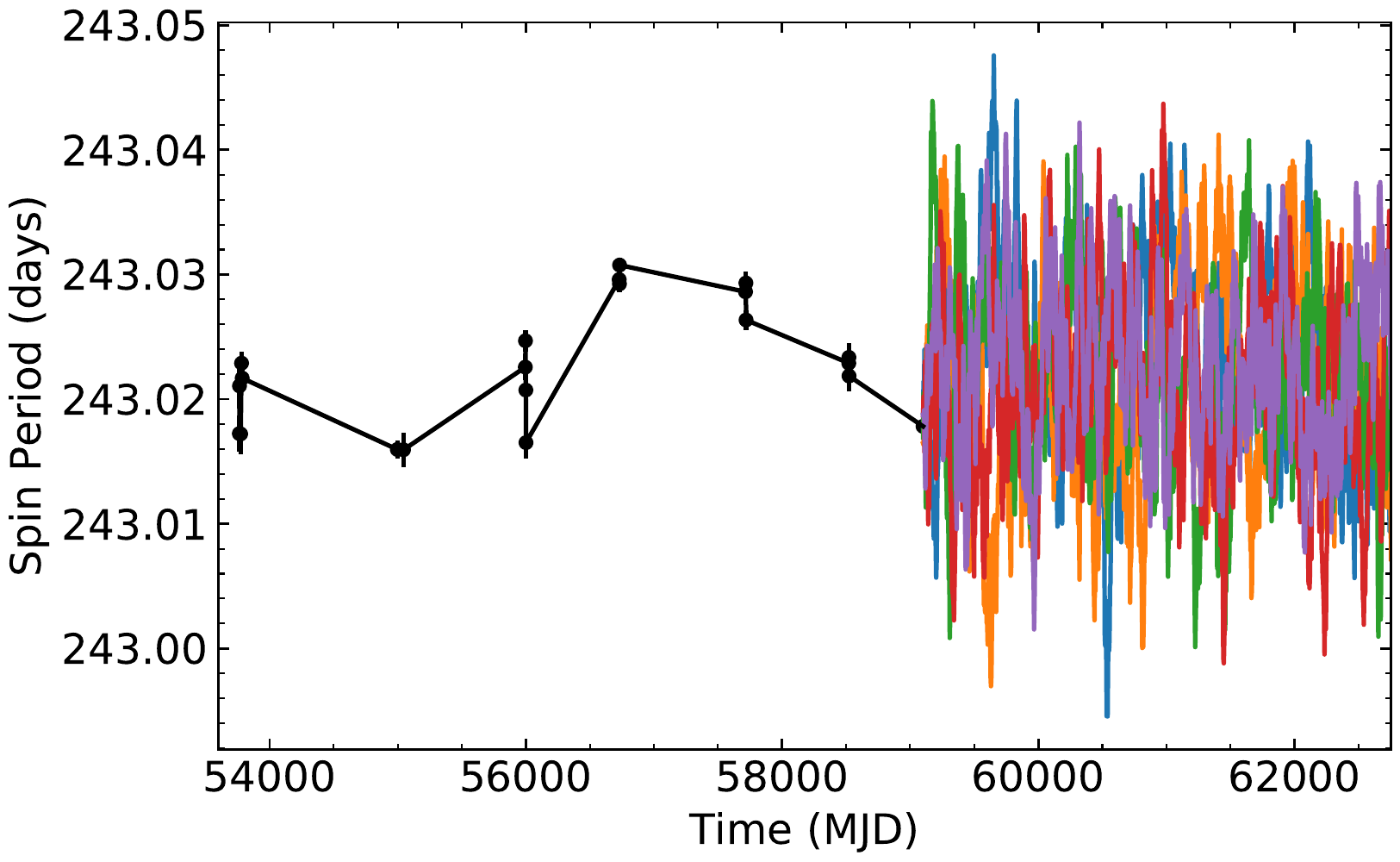} 
  \caption{ (Top) Prediction of the evolution of the spin period of
    Venus over 10 years with a single realization (blue line) of a
    continuous autoregressive moving average (CARMA) model of order
    $(p,q)=(1,0)$ whose coefficients have been adjusted to provide a
    good fit to the existing data (black points with error bars).
    (Bottom) Predictions for five distinct realizations (colored
    lines).}
    \label{fig-carma}
\end{figure}

\begin{figure}[p]
  \centering
  \includegraphics[angle=0,width=4.7in]{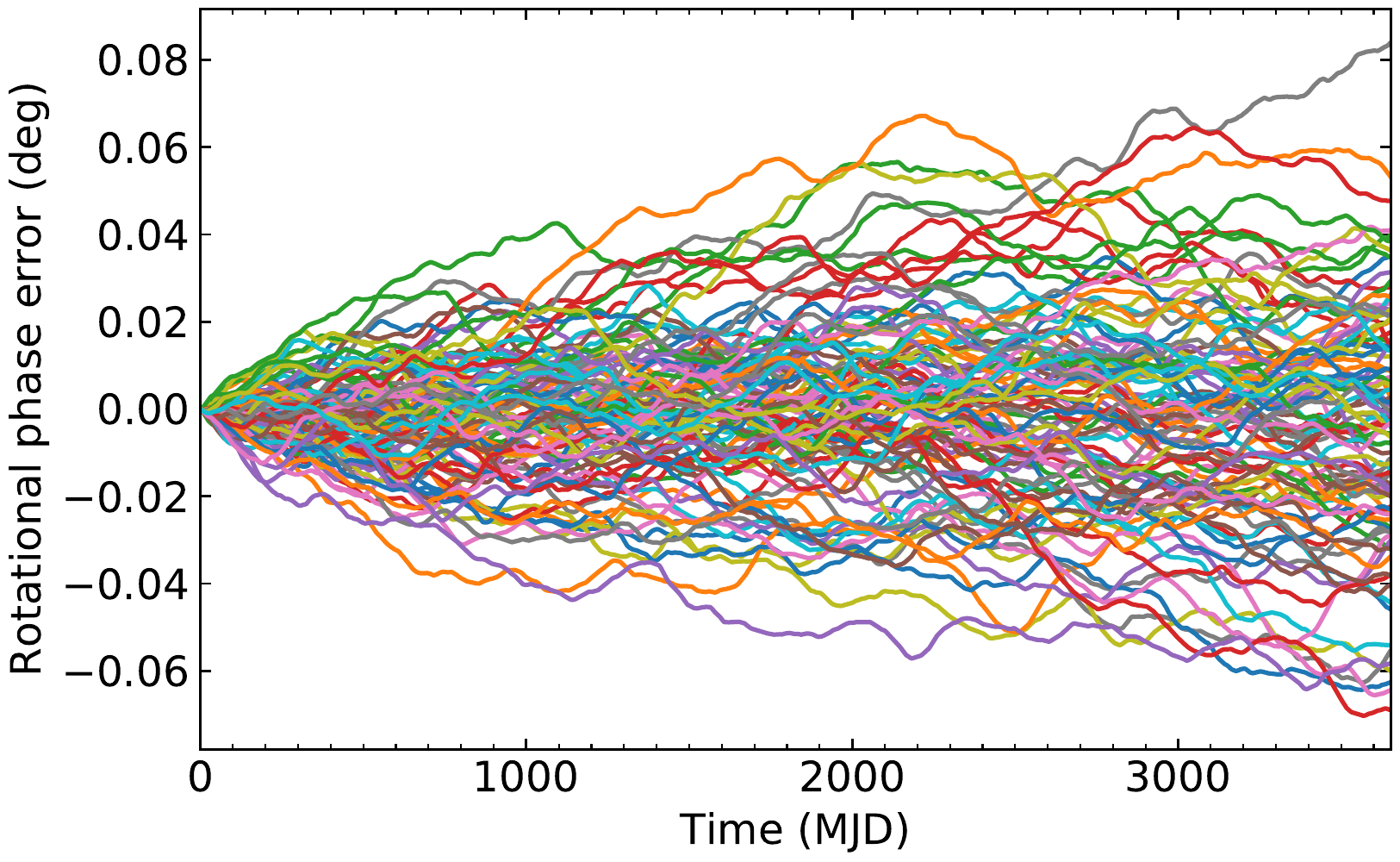} 
  \caption{Ten-year predictions of the evolution of the rotational
    phase difference between a model of Venus rotating with a variable
    spin period and a model of Venus rotating with a constant spin
    period ($P=243.0226$ days).  One hundred distinct realizations of
    a CARMA$(p,q)$ process with $(p,q)=(1,0)$ are shown.
      Because the
model has a random walk component, the rotational phase errors
increase approximately as the square root of the forecast time interval.
An error of    0.02 deg corresponds to 2.1~km at the equator.  
}
    \label{fig-evolution}
\end{figure}

We followed the methods of Kelly et al.\ \cite{kell14}.  First we chose
the $(p,q)$ order of the CARMA model that provided the best
representation of our data.  We tested all values of $p < 5, q < p$
and found that $(p,q)=(0,1)$ yielded the best results according to an
Akaike information criterion with correction for finite sample sizes
(AICc).  This model is also known as a Gaussian first-order
continuous-time autoregressive process (CAR(1)) or stationary Gaussian
Ornstein-Uhlenbeck process.  We used a Bayesian framework to optimize
model parameters, which included a Markov Chain Monte Carlo technique
to generate the posterior probability distributions of the model
parameters.  The model includes a scaling parameter on the measurement
errors to capture possible
under-estimation of the assigned errors.  In practice, the model
increased the measurement error variance by $\sim$7\%.  Once the model
was optimized, we generated independent realizations of the time
evolution of the spin period with time increments of one day over a
period of 10 years (Supplementary Fig.~\ref{fig-carma}).  We computed
time integrals to obtain rotational phase histories and subtracted the
rotational phase history of a model with constant spin period
(Supplementary Fig.~\ref{fig-evolution}).  Finally, we evaluated
the spread in the rotational phase error observed in 2000 distinct
realizations to assess the likely error in inertial positions after 10
years.

We found that LOD variations on Venus render the predictions of the
absolute positions of surface locations uncertain by $\pm$3.3~km
(1$\sigma$) after ten years.  Because the model has a random walk
component, the rotational phase errors increase approximately as the
square root of the forecast time interval.  The stochastic LOD
variations will complicate the establishment of new geodetic networks,
which typically rely on rotational models with uniform spin period or
simple harmonic expressions for the spin period.  Likewise, the
rotational noise will challenge lander-based or landmark-based
attempts at measuring the spin precession rate because the relatively small
systematic changes in the inertial positions of the landers or
landmarks due to precession will be polluted by the much larger
stochastic noise due to AAM variations.

For instance, a future lander that would survive on the surface for 60
days would experience a total displacement of $<$10~m due to the
precession, whereas the stochastic rotational noise over the mission
duration would be $\sim$420~m.  Achieving a 5\% precision on the spin
precession rate would require a measurement at the 0.5~m level.

A future orbiter with 5-year mission duration would have to measure a
$\sim$300~m displacement of landmarks due to precession with a
precision of 15~m to determine the precession rate with 5\%
uncertainties.  This measurement would have to content with a
background rotational noise of $\sim$2300~m over the 5-year period.

\begin{figure}[p]
  \centering
  \includegraphics[angle=0,width=4.3in]{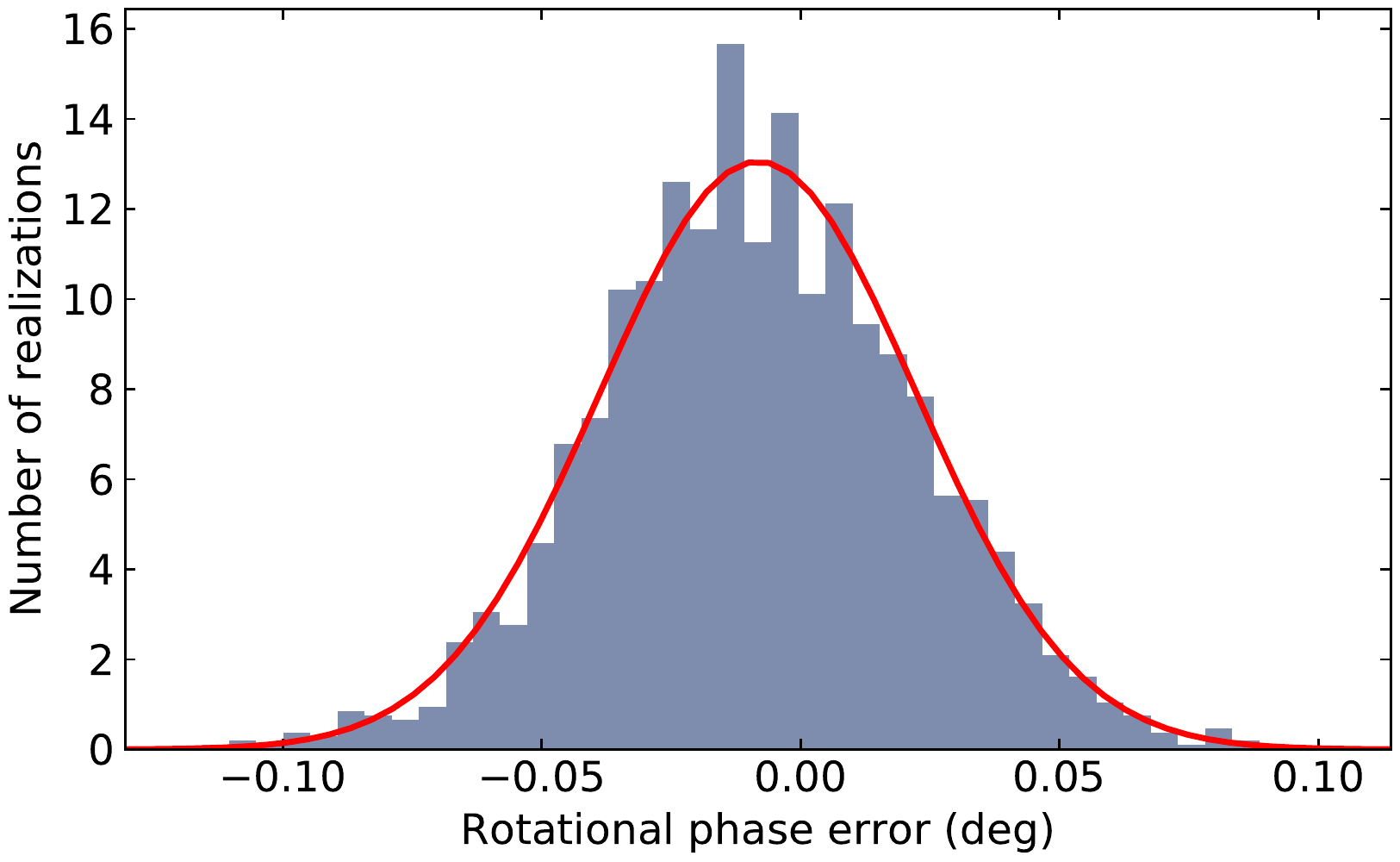} 
  \caption{Histogram of the rotational phase discrepancies accumulated
    after 10 years between a variable spin model and a constant spin
    model ($P=243.0226$ days).  The results of 2000 distinct
    realizations of a CARMA$(p,q)$ process with $(p,q)=(1,0)$ are
    shown.  The standard deviation of the Gaussian is 0.031 deg, which
    amounts to a $\pm$3.3~km prediction error at the equator.
    The ensemble average of the rotational phase error is not zero
    because the coefficients of the CARMA model were sampled from
    posterior distributions and resulted in a model that favor a
    slightly different average spin period than our adopted average spin period.
  }
    \label{fig-hist2000}
\end{figure}

\section{Identification of candidate periodicities with Lomb periodograms and phase dispersion minimization analyses}

\begin{figure}[H]
  \centering
  \includegraphics[angle=0,width=4in]{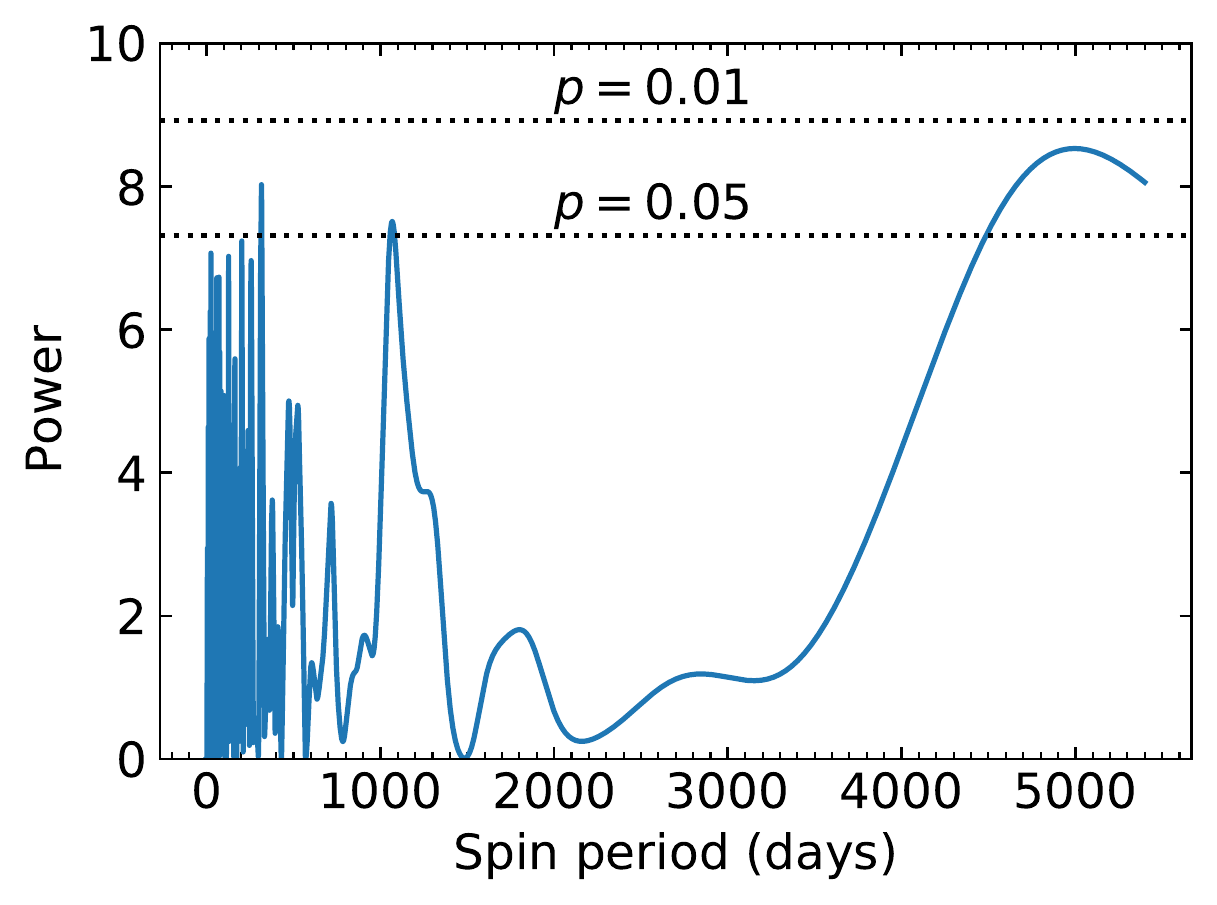} 
  \includegraphics[angle=0,width=4in]{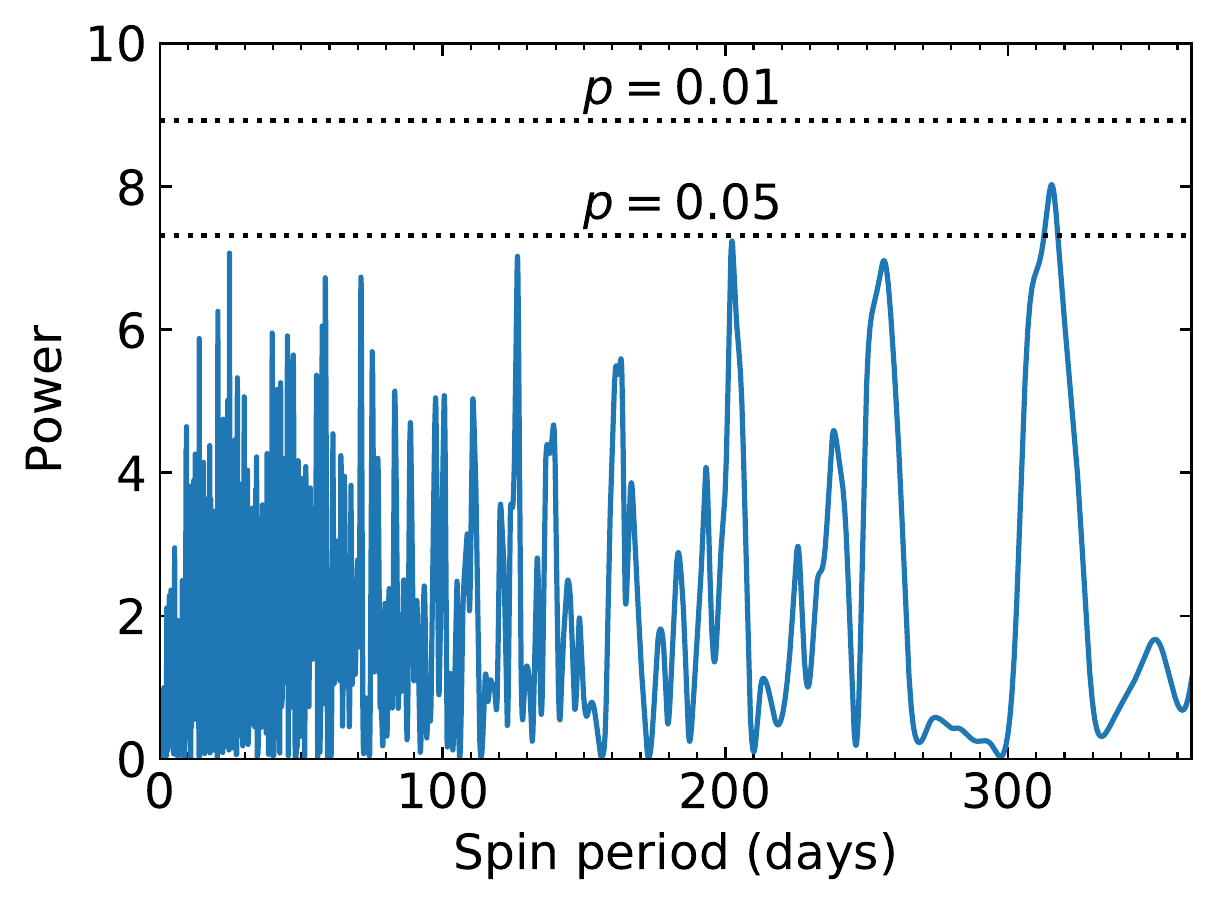} 
    \caption{ (Top) Lomb periodogram analysis for trial periods between 1 and 5400
      days computed with increments of 0.1 days.  The horizontal
      dashed lines show the false alarm probabilities or significance
      levels.  (Bottom) Same periodogram showing periods between 1 and 365 days.}
    \label{fig-Lomb5400}
\end{figure}

\begin{table}[H]
\begin{tabular}{rrr}
  $P$ (days) & Power & $p$-value\\
  \hline
4996.40 & 8.5324 & 1.48e-02\\
315.50 & 8.0274 & 2.45e-02\\
1067.90 & 7.5130 & 4.09e-02\\
202.40 & 7.2402 & 5.38e-02\\
24.60 & 7.0704 & 6.37e-02\\
126.50 & 7.0256 & 6.67e-02\\
256.10 & 6.9658 & 7.08e-02\\
71.10 & 6.7347 & 8.92e-02\\
58.50 & 6.7261 & 8.99e-02\\
39.70 & 5.9531 & 1.95e-01\\
\end{tabular}
\caption{Top 10 results from Lomb periodogram analysis of Venus LOD
  variations, with trial periods ranging from 1 day to 5400 days in
  increments of 0.1 days. Columns show trial period, power, and the
  probability of false alarm or significance level.}
\label{tab-Lomb5400}
\end{table}

\begin{table}[H]
\begin{tabular}{rrr}
  $P$ (days) & $\theta_{\rm PDM}$ & $p$-value\\
  \hline
47.91 & 0.0114  & 0.00e+00\\
12.71 & 0.0121  & 1.45e-02\\
13.10 & 0.0132  & 2.30e-02\\
18.34 & 0.0173  & 1.75e-02\\
 1.25 & 0.0181  & 2.05e-02\\
75.07 & 0.0221  & 1.50e-03\\
62.28 & 0.0238  & 5.00e-04\\
15.96 & 0.0242  & 3.35e-02\\
32.66 & 0.0307  & 1.40e-02\\
14.23 & 0.0343  & 4.90e-02\\
\end{tabular}
\caption{Top 10 results from PDM analysis of Venus LOD variations,
  with trial periods ranging from 1 day to 5400 days in increments of
  0.01 days. Columns show trial period, PDM statistic $\theta$, and a
  pseudo $p$-value obtained by ranking 2000 Monte Carlo trials where
  the observed spin periods are permuted prior to PDM analysis.}
\label{tab-PDM}
\end{table}

\begin{figure}[H]
  \centering
  \includegraphics[angle=0,width=3.1in]{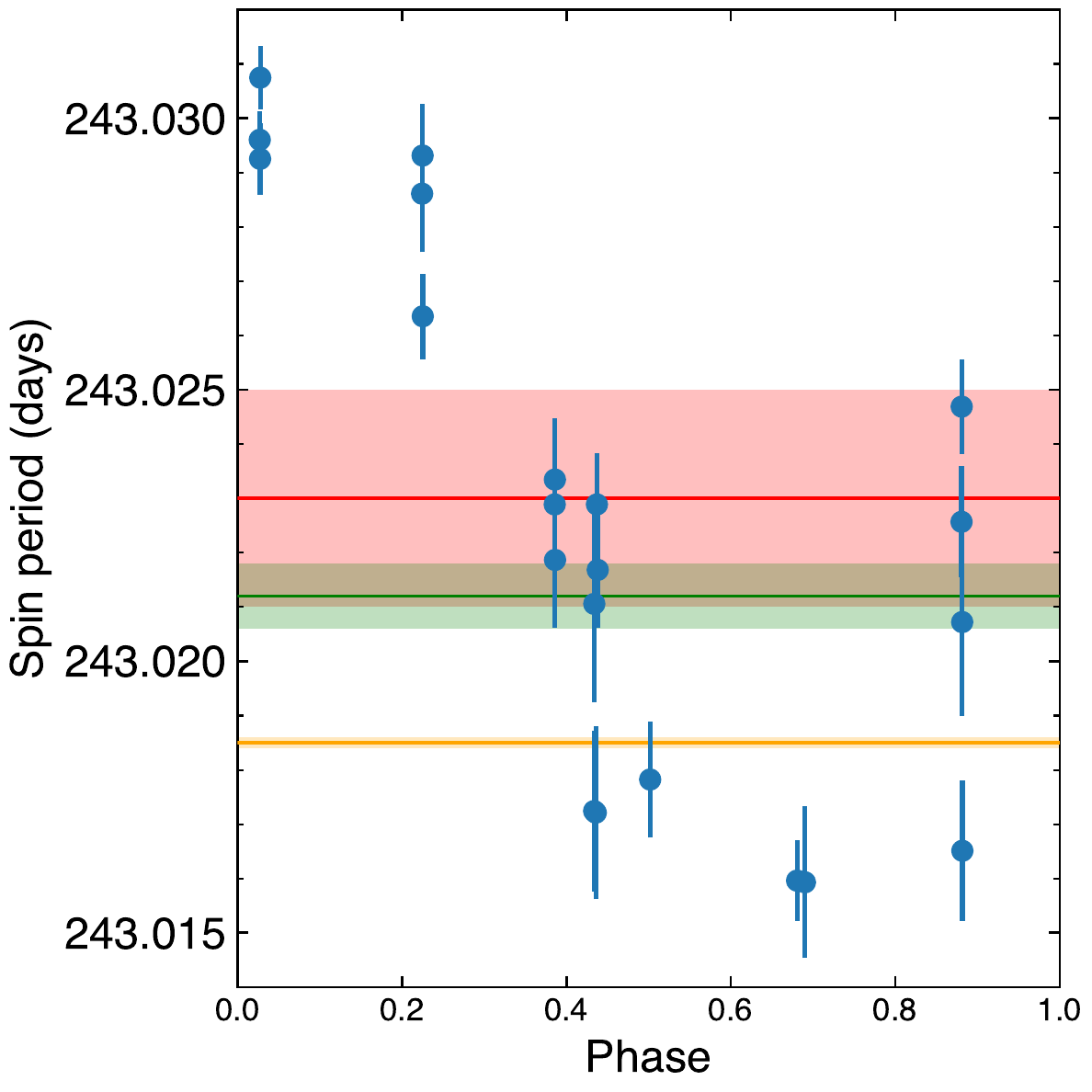} 
  \includegraphics[angle=0,width=3.1in]{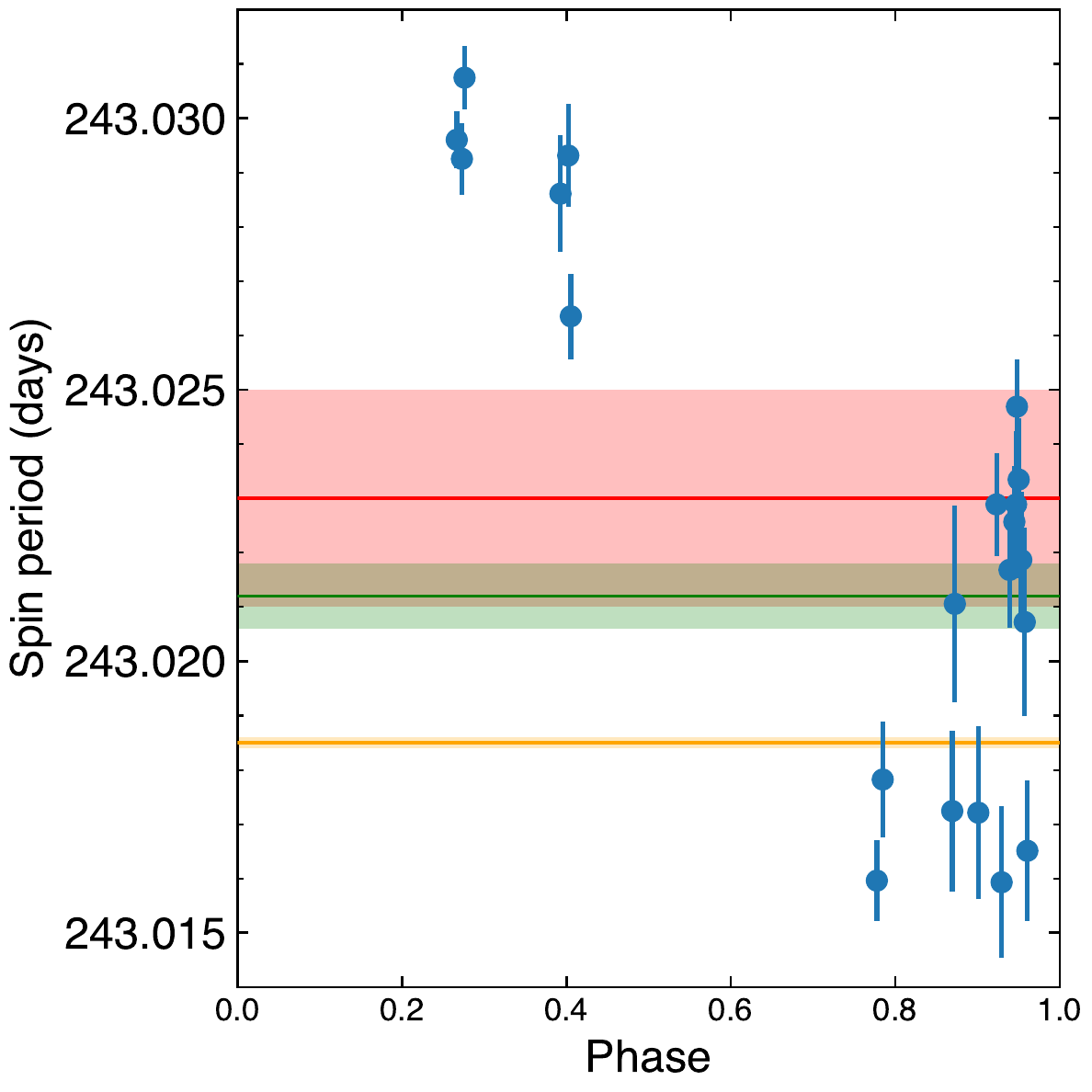} 
  \includegraphics[angle=0,width=3.1in]{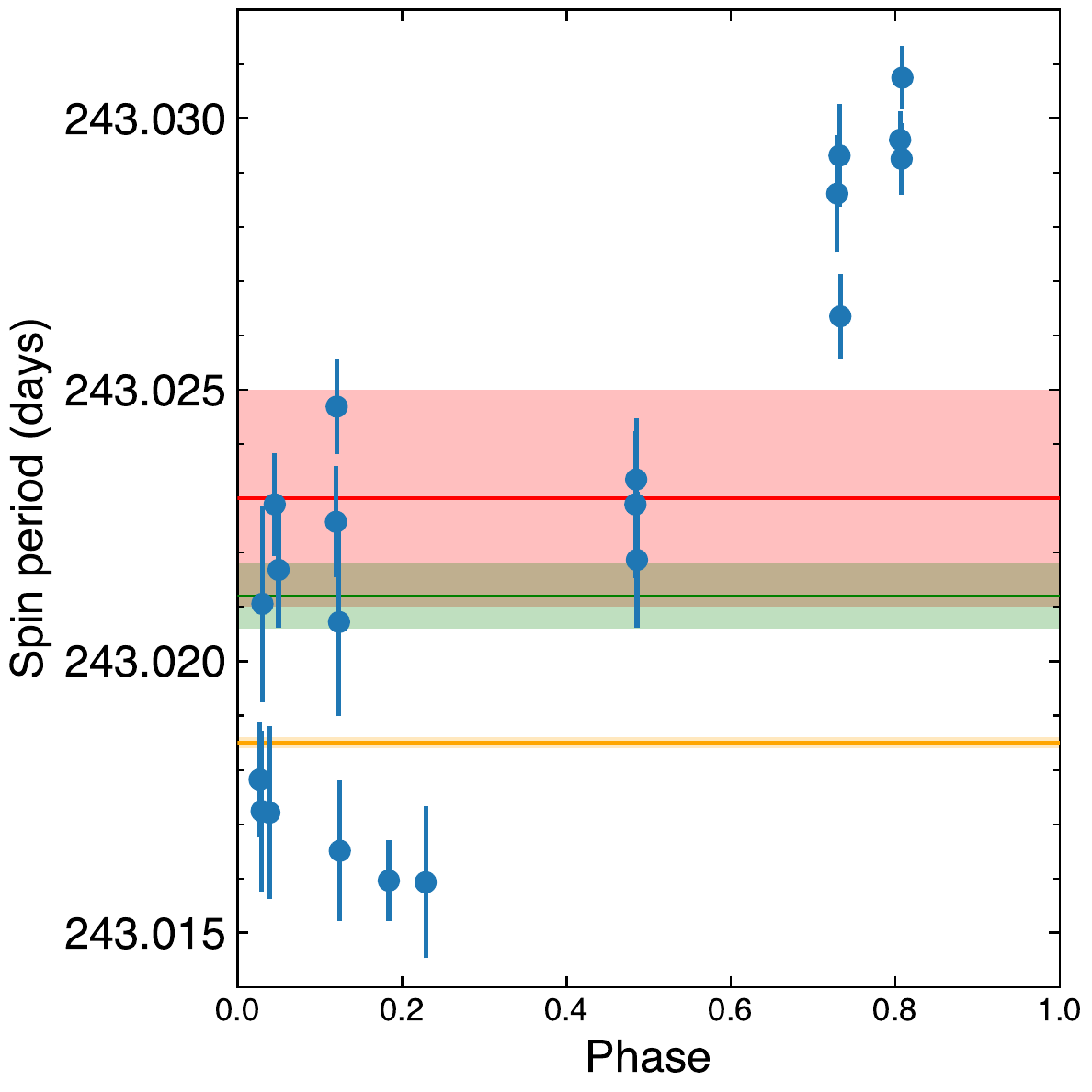}
  \includegraphics[angle=0,width=3.1in]{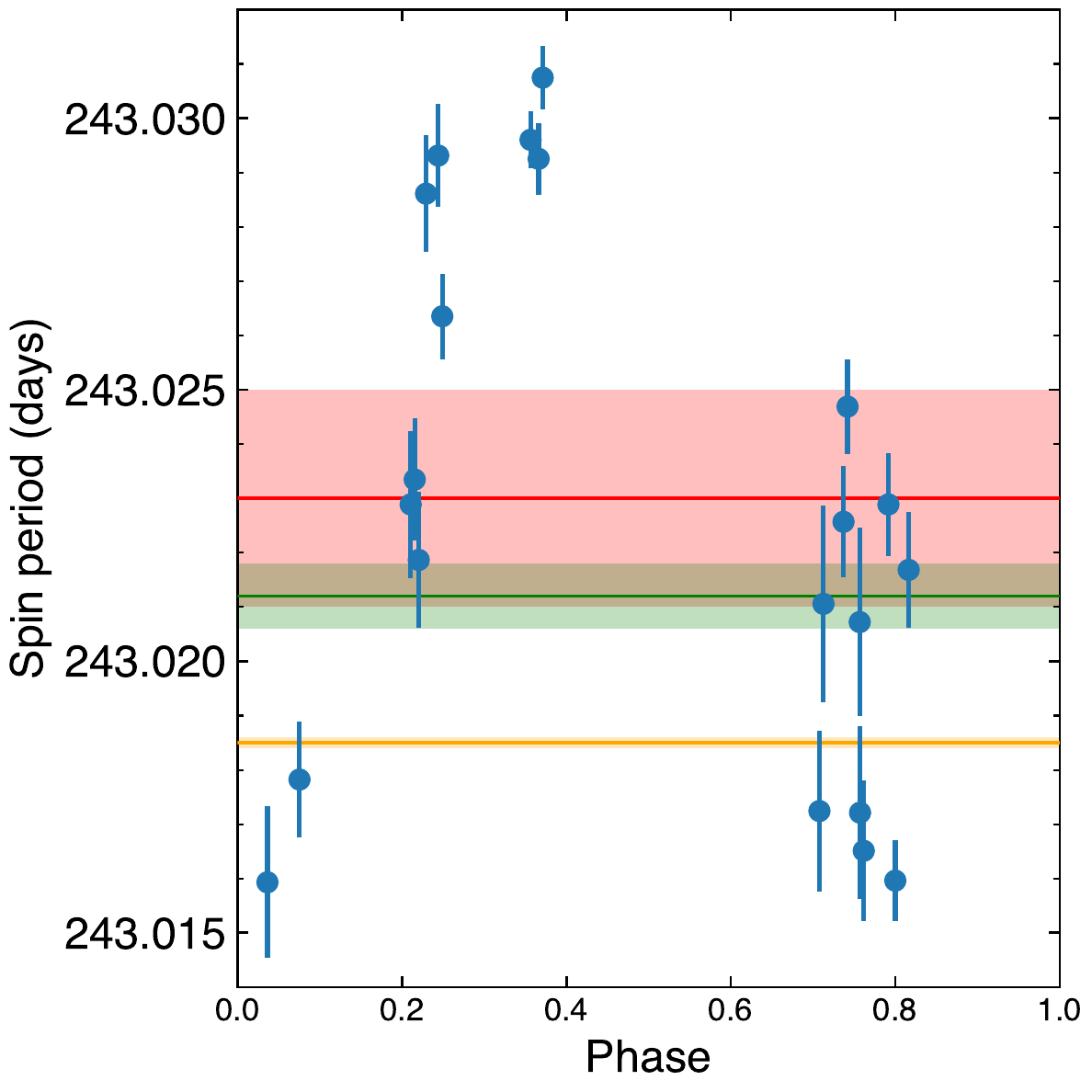}
    \caption{ Instantaneous spin period measurements of Venus folded
      onto candidate periodicities at ({\bf Top left}) 4996.4 days, 
      ({\bf Top right}) 315.5 days, ({\bf Bottom left}) 1067.9 days, 
      and ({\bf Bottom right}) 202.4 days.  The origin of phase is 2000 FEB 22
      10:43:58 TDB.}
    \label{fig-top4}
\end{figure}
\vfill\mbox{}

\section{Identification of candidate periodicities related to spin, orbital, and diurnal frequencies}

\begin{figure}[H]
  \centering
  \includegraphics[angle=0,width=4in]{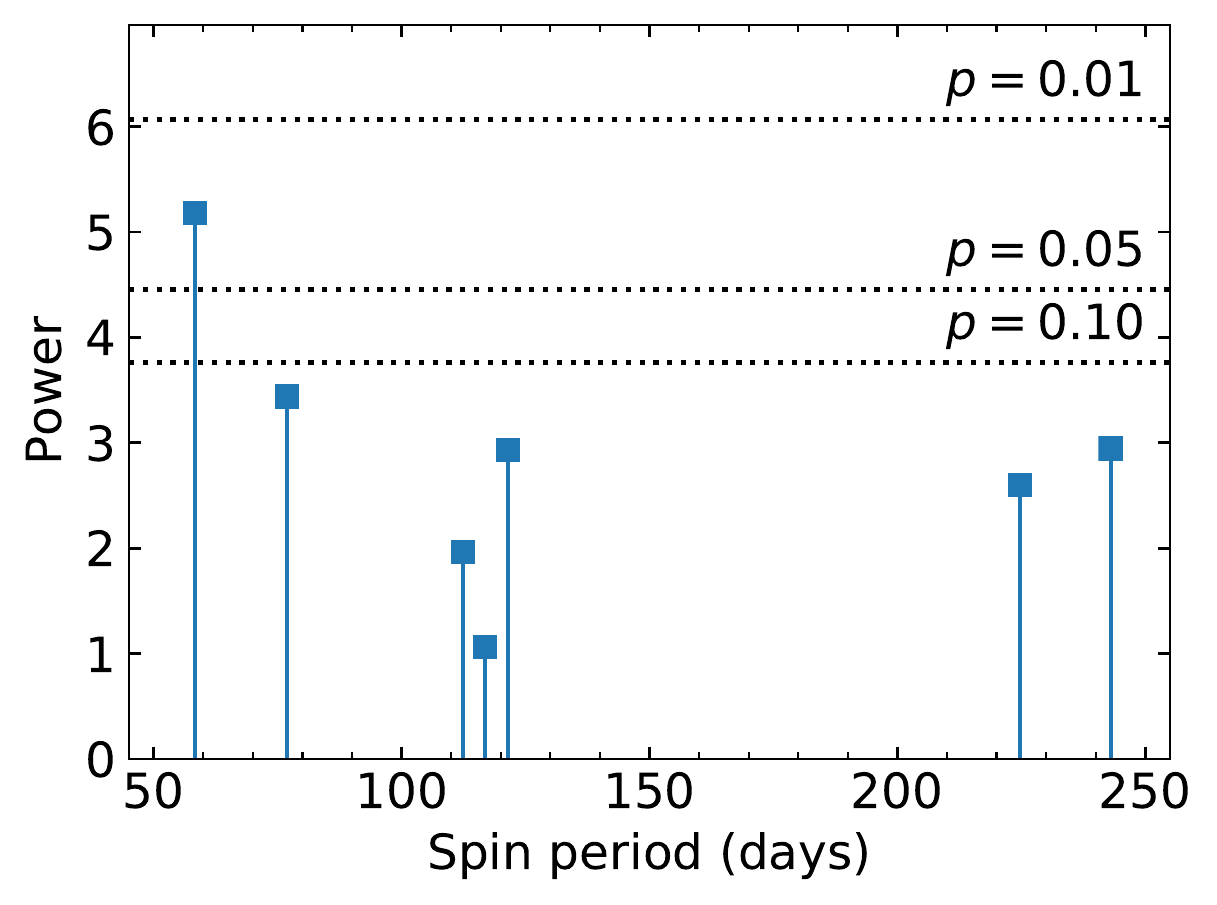} 
    \caption{ Lomb periodogram analysis of periodicities related to the to the spin
      ($\omega$), orbital ($n$), and diurnal ($\omega + n$)
      frequencies.  The horizontal dashed lines show the false alarm
      probabilities or significance levels.  Only the semidiurnal
      period at 58.4 days reaches a $p=0.05$ level of significance by this metric.}
    \label{fig-Lomb}
\end{figure}

\begin{table}[H]
\begin{tabular}{rrrr}
$P$ (days) & Power & $p$-value & Frequency\\
\hline
243.023  & 2.9454 & 2.26e-01 & $\omega$      \\
224.701 & 2.5981 & 3.20e-01  & $n$           \\
121.511 & 2.9294 & 2.30e-01  & $2\omega$     \\
116.752 & 1.0657 & 1.48e+00  & $\omega + n$  \\
112.350 & 1.9627 & 6.04e-01  & $2n$          \\
76.831 & 3.4380 & 1.38e-01   & $\omega + 2n$ \\
58.376 & 5.1808 & 2.42e-02   & $2\omega+ 2n$ \\
\end{tabular}
\caption{Lomb periodogram analysis of periodicities in Venus LOD
  variations, with trial periods associated with spin ($\omega$),
  orbital ($n$), and diurnal ($\omega + n$) frequencies. Columns show
  trial period, power, the probability of false alarm or significance
  level, and frequency.}
\label{tab-Lombphys}
\end{table}

\begin{table}[H]
\begin{tabular}{rrrr}
$P$ (days) & $\theta_{\rm PDM}$ & $p$-value & Frequency\\
\hline
243.023  &  0.5811   &  8.70e-02& $\omega$      \\
224.701  &  0.2722   &  1.50e-03& $n$           \\
121.511  &  0.6714   &  2.07e-01& $2\omega$     \\
116.752  &  0.1856   &  5.00e-04& $\omega + n$  \\
112.350  &  0.4430   &  4.00e-02& $2n$          \\
 76.831  &  0.6204   &  3.04e-01& $\omega + 2n$ \\
 58.376  &  0.1842   &  4.20e-02& $2\omega+ 2n$ \\
\end{tabular}
\caption{Results of PDM analysis of
  Venus LOD variations, with
  trial periods associated with spin
      ($\omega$), orbital ($n$), and diurnal ($\omega + n$)
  frequencies. Columns show trial period, PDM statistic $\theta$, a
  pseudo $p$-value obtained by ranking 2000 Monte Carlo trials where
  the observed spin periods are permuted prior to PDM analysis, and
  frequency.  The semidiurnal period ranks first, but only marginally
  so compared to the diurnal period.  The orbital period is a distant
  third.}
\label{tab-PDMphys}
\end{table}

\pagebreak

\clearpage



\end{document}